\documentstyle[epsfig]{mn2e}
\begin{document}

\title[Cosmological baryonic and matter densities from SDSS
  LRGs]{Cosmological baryonic and matter densities from 600{,}000 SDSS
  Luminous Red Galaxies with photometric redshifts}

\author[Chris Blake et al.]{Chris Blake$^{1,}$\footnotemark, Adrian
  Collister$^2$, Sarah Bridle$^3$, Ofer Lahav$^3$ \\ \\ $^1$
  Department of Physics \& Astronomy, University of British Columbia,
  6224 Agricultural Road, Vancouver, B.C., V6T 1Z1, Canada \\ $^2$
  Institute of Astronomy, University of Cambridge, Cambridge, CB3 0HA,
  U.K. \\ $^3$ Department of Physics \& Astronomy, University College
  London, Gower Street, London, WC1E 6BT, U.K.}

\maketitle

\begin{abstract}
We analyze MegaZ-LRG, a photometric-redshift catalogue of Luminous Red
Galaxies (LRGs) based on the imaging data of the Sloan Digital Sky
Survey (SDSS) 4th Data Release.  MegaZ-LRG, presented in a companion
paper, contains $> 10^6$ photometric redshifts derived with ANNz, an
Artificial Neural Network method, constrained by a spectroscopic
sub-sample of $\approx 13{,}000$ galaxies obtained by the 2dF-SDSS LRG
and Quasar (2SLAQ) survey.  The catalogue spans the redshift range
$0.4 < z < 0.7$ with an r.m.s.\ redshift error $\sigma_z \approx
0.03(1+z)$, covering $5{,}914$ deg$^2$ to map out a total cosmic
volume $2.5 \, h^{-3}$ Gpc$^3$.  In this study we use the most
reliable $600{,}000$ photometric redshifts to measure the large-scale
structure using two methods: (1) a spherical harmonic analysis in
redshift slices, and (2) a direct re-construction of the spatial
clustering pattern using Fourier techniques.  We present the first
cosmological parameter fits to galaxy angular power spectra from a
{\it photometric} redshift survey.  Combining the redshift slices with
appropriate covariances, we determine best-fitting values for the
matter density $\Omega_{\rm m}$ and baryon density $\Omega_{\rm b}$ of
$\Omega_{\rm m} h = 0.195 \pm 0.023$ and $\Omega_{\rm b}/\Omega_{\rm
  m} = 0.16 \pm 0.036$ (with the Hubble parameter $h = 0.75$ and
scalar index of primordial fluctuations $n_{\rm scalar} = 1$ held
fixed).  These results are in agreement with and independent of the
latest studies of the Cosmic Microwave Background radiation, and their
precision is comparable to analyses of contemporary
spectroscopic-redshift surveys.  We perform an extensive series of
tests which conclude that our power spectrum measurements are robust
against potential systematic photometric errors in the catalogue.  We
conclude that photometric-redshift surveys are competitive with
spectroscopic surveys for measuring cosmological parameters in the
simplest ``vanilla'' models.  Future deep imaging surveys have great
potential for further improvement, provided that systematic errors can
be controlled.
\end{abstract}

\section{Introduction}
\renewcommand{\thefootnote}{\fnsymbol{footnote}}
\setcounter{footnote}{1}
\footnotetext{E-mail: cab@astro.ubc.ca}

The large-scale structure of the Universe is one of the most important
probes of cosmology, encoding information about the fundamental
parameters of the Universe and the physical processes governing the
development of cosmic structure and the formation of galaxies.  In
particular, the large-scale power spectrum of matter fluctuations
depends on a different combination of cosmological parameters to the
Cosmic Microwave Background anisotropy spectrum.  The combination of
these two datasets provided the first compelling evidence for the
current standard model of cosmology, in which two-thirds of the
present energy density of the Universe is resident in some form of
``dark energy'' which is driving the acceleration of the cosmic
expansion rate (e.g.\ Efstathiou, Sutherland \& Maddox 1990; Ostriker
\& Steinhardt 1995).  This model was spectacularly confirmed by
subsequent observations of distant supernovae (e.g.\ Riess et
al.\ 1998; Perlmutter et al.\ 1999).  Measurements of large-scale
structure have continued to constitute an important component of
cosmological parameter fits (e.g.\ Efstathiou et al.\ 2002; Percival
et al.\ 2002; Spergel et al.\ 2003; Pope et al.\ 2004; Tegmark et
al.\ 2004b; Seljak et al.\ 2005; Sanchez et al.\ 2006).

The pattern of cosmic structure has been delineated by a series of
increasingly impressive galaxy redshift surveys.  The current
state-of-the-art is represented by the 2-degree Field Galaxy Redshift
Survey (2dFGRS; Colless et al.\ 2001) and the Sloan Digital Sky Survey
(SDSS; York et al.\ 2000), which have provided exquisitely detailed
maps of the local ($z \sim 0.1$) Universe (e.g.\ Percival et
al.\ 2001; Cole et al.\ 2005; Tegmark et al.\ 2004a; Eisenstein et
al.\ 2005).  At higher redshifts ($z \sim 1$), surveys by the Deep
Extragalactic Evolutionary Probe (DEEP2; Davis et al.\ 2003) and the
VIRMOS-VLT Deep Survey (VVDS; Le Fevre et al.\ 2003) have produced
impressive measurements of the clustering pattern over an area of a
few square degrees (e.g.\ Coil et al.\ 2004; Le Fevre et al.\ 2005).
But despite these efforts, today's most pressing cosmological problems
-- such as the investigation of the nature of dark energy, the
discrimination between competing models of inflation, and the
measurement of neutrino masses -- demand the construction of vastly
larger spectroscopic surveys of millions of faint galaxies spanning
thousands of degrees at high redshift.

The pattern of large-scale structure can also be measured by imaging
surveys via its projection on the sky.  Although the lack of redshift
information weakens the precision with which the clustering functions
can be determined, this is partially compensated for by the ease with
which large areas of sky can be surveyed.  For example, the APM Galaxy
Survey, constructed by scanning photographic plates covering several
thousand square degrees of the southern galactic cap, was used in the
early 1990s to produce the best then-existing measurements of the {\it
  three-dimensional} large-scale clustering pattern (Baugh \&
Efstathiou 1993).

Moreover, the power of imaging surveys can be greatly enhanced by the
use of {\it photometric redshifts}, which are estimated from broadband
galaxy colours rather than from spectra.  These redshift estimates
enable the catalogued galaxies to be divided into many
quasi-independent radial slices, greatly improving the signal-to-noise
ratio of the resulting clustering measurements.  The utility of
photometric redshifts is now well-established, with many successful
techniques being employed.  The r.m.s. precision $\sigma_z$ with which
galaxy redshifts may be determined varies with the method and filter
set used, together with the galaxy type, magnitude and redshift, but
at best is currently $\sigma_0 \equiv \sigma_z / (1 + z) \sim 0.03$
(e.g.\ COMBO-17; Wolf et al.\ 2003).  Photometric redshifts have
already been used to construct volume-limited samples of low-redshift
galaxies, and measure their angular clustering properties as a
function of luminosity and rest-frame colour (e.g.\ Budavari et
al.\ 2003).

Certain sub-classes of galaxy provide especially accurate photometric
redshifts, in particular {\it Luminous Red Galaxies} (LRGs), for which
the optical colours change very rapidly with redshift owing to a
significant spectral break at 4000\AA\, (see Eisenstein et al.\ 2001;
Padmanabhan et al.\ 2005).  Moreover, these galaxies inhabit massive
dark matter haloes (Zehavi et al.\ 2005) and are consequently highly
efficient (although biased) tracers of the underlying clustering
pattern.  In this sense, a sub-sample of LRGs should provide the
optimal set of tracers of the large-scale structure.  Although by
constructing such a sub-sample we are excluding a large fraction of
the overall galaxy population, this is immaterial (as far as
determining cosmological parameters is concerned) provided that the
survey is sufficiently deep that the contribution of shot noise to the
error in the clustering measurements is negligible.

In this study we perform a clustering analysis on a large sample of
$\sim 10^6$ Luminous Red Galaxies selected photometrically from the
imaging component of the SDSS 4th Data Release (DR4), spanning the
redshift range $0.4 < z < 0.7$.  We believe that this investigation is
timely for several reasons.  Firstly, we aim to demonstrate the
feasibility of extracting high-quality large-scale structure
measurements from an imaging dataset employing photometric redshifts.
Several forthcoming imaging surveys are planning to produce accurate
cosmological measurements in this manner -- for example, the
Kilo-Degree Survey (KIDS), the Dark Energy Survey (DES), and the
Panoramic Survey Telescope (PanStarrs) -- whereas next-generation
spectroscopic surveys still require the development of challenging new
instrumentation such as the Wide-Field Multi-Object Spectograph
(WFMOS) or the Square Kilometre Array (SKA).  The cosmological
parameter constraints resulting from future photometric redshift
imaging surveys have been simulated by Seo \& Eisenstein (2003);
Amendola, Quercellini \& Giallongo (2004); Dolney, Jain \& Takada
(2004); Blake \& Bridle (2005) and Zhan et al.\ (2006).

Secondly, the recent detection of {\it baryon acoustic oscillations}
in the galaxy clustering pattern (Eisenstein et al.\ 2005; Cole et
al.\ 2005; Huetsi 2006a) has provided a powerful new cosmological
probe.  The existence of this preferred scale or standard ruler in the
galaxy distribution will eventually permit accurate measurements of
the properties of dark energy (Blake \& Glazebrook 2003, Seo \&
Eisenstein 2003).  Eisenstein et al.\ (2005) and Huetsi (2006a)
identified the preferred acoustic scale in the distribution of LRGs in
the SDSS {\it spectroscopic} sample, with mean redshift $\overline{z}
\sim 0.35$.  However, the baryon oscillation scale can in principle be
extracted from a photometric-redshift survey, and the completed SDSS
imaging survey could produce a significant detection at an
interestingly higher effective redshift of $\overline{z} \sim 0.55$
(Blake \& Bridle 2005).

Thirdly, this dataset provides a useful test case for the comparison
of different methods for measuring the large-scale structure.  In this
study we consider two techniques: the determination of the
(two-dimensional) angular power spectrum in redshift slices via a
spherical harmonic analysis (Peebles 1973); and a direct
re-construction of the (three-dimensional) spatial power spectrum
using a Fourier analysis which only retains large-scale radial
components (Seo \& Eisenstein 2003).  The direct Fourier treatment may
be preferable for the detection of features in the power spectrum
because it avoids unnecessary binning of data along the line-of-sight
(which is equivalent to a convolution or smoothing of the measured
power spectrum).  However, the direct measurement of the spatial power
spectrum requires an assumption of values for the cosmological
parameters, thus parameter-fitting becomes a more complex iterative
exercise.

Finally, this survey provides an ideal demonstration of the latest
techniques for determining photometric redshifts.  We use ANNz, an
artificial neural network approach (Firth, Lahav \& Somerville 2003)
which compares very well with other successful photometric redshift
methods (see Collister \& Lahav 2004).  The neural network technique
requires a ``training set'' of spectroscopic redshift measurements for
a sub-sample of the photometric catalogue.  We use data from the
2dF-SDSS LRG and Quasar (2SLAQ) survey for this purpose (Cannon et
al.\ 2006).  The resulting photo-$z$ catalogue of $\sim 10^6$ LRGs,
which we have named MegaZ-LRG, is described by Collister et
al.\ (2006).

This paper is organized as follows.  In Section \ref{secdata} we
describe the construction of the LRG catalogue from publicly-available
SDSS data, the definition of the angular selection function, and the
determination and validation of the photometric redshifts and error
distribution.  Section \ref{secangpow} details the measurement of the
angular power spectrum in redshift slices using a spherical harmonic
analysis, the fitting of cosmological models and the resulting
probability contours for the cosmological parameters.  Section
\ref{secsys} contains an extensive series of tests for systematic
photometric errors in the imaging data that may potentially influence
these clustering measurements.  Section \ref{secspatpow} presents a
direct determination of the spatial power spectrum via a Fourier
analysis and demonstrates consistency with the angular power spectrum
results.  We summarize our conclusions in Section \ref{secconc}.

We note that a parallel investigation of a similar dataset was carried
out by Padmanabhan et al.\ (2006) using a different
photometric-redshift method and power spectrum estimation procedure.
Where our results can be compared, the agreement is good.

\section{Data}
\label{secdata}

We analyze galaxy clustering in the MegaZ-LRG database, a
photometric-redshift catalogue of Luminous Red Galaxies based on the
imaging component of the SDSS 4th Data Release.  The construction of
this catalogue is described in detail by Collister et al.\ (2006), and
we only provide a brief description here.

\subsection{Selection criteria}

We selected Luminous Red Galaxies at intermediate redshifts ($0.4 < z
< 0.7$) from the SDSS imaging database using a series of colour and
magnitude cuts (Collister et al.\ 2006).  These cuts were designed to
match the selection criteria of the 2dF-SDSS LRG and Quasar (2SLAQ)
survey (Cannon et al.\ 2006), a spectroscopic follow-up to the SDSS
imaging survey using the 2-degree Field (2dF) spectrograph at the
Anglo-Australian Telescope.  These 2SLAQ spectroscopic redshifts were
used to train and test the photometric redshift code, which we then
applied to the entire set of LRGs selected from the SDSS imaging
database.  A similar procedure was carried out for earlier versions of
the SDSS and 2SLAQ databases by Padmanabhan et al.\ (2005).

The 2SLAQ survey obtained good-quality redshifts for about $13{,}000$
objects in selected fields of the SDSS equatorial stripe (at
declination $\delta \approx 0^\circ$).  Applying the same 2SLAQ
selection criteria to the entire SDSS DR4 imaging catalogue returned
$\sim 10^6$ LRGs.  The 2SLAQ survey demonstrated that these selection
criteria are $\approx 95\%$ efficient in the identification of
intermediate-redshift LRGs.  The most significant contaminant,
accounting for virtually all of the remaining $\approx 5\%$ of
objects, is M-type stars.  In the 2SLAQ survey these stars, which have
very similar $gri$ colours to the LRGs, are readily excluded using
spectra.  This is not possible for a photometrically-selected sample,
but star-galaxy separation cuts can be used to reduce the stellar
fraction further, as described below.

The 2SLAQ selection criteria fluctuated a little at the beginning of
the survey.  Specifically, the faint limit of the $i$-band magnitude
$i_{\rm deV}$, and the minimum value of $d_{\rm perp}$ (a colour
variable used to select LRGs), were varied slightly.  For the majority
of the 2SLAQ survey, the criteria $i_{\rm deV} \le 19.8$ and $d_{\rm
  perp} \ge 0.55$ were used.  In this paper we analyze a conservative
version of the photometric-redshift catalogue in which these limits
are strictly enforced, avoiding any extrapolation of the ``training
set'' calibration.

We note that our training sub-sample is extrapolated in sky position.
The 2SLAQ targets lie exclusively in the equatorial stripe at
declination $\delta \approx 0^\circ$, so may not fully trace effects
such as dust extinction which depend on sky co-ordinate.  However, we
find our results to be independent of varying dust extinction, as
described in Section \ref{secsys}.

\begin{figure*}
\center
\epsfig{file=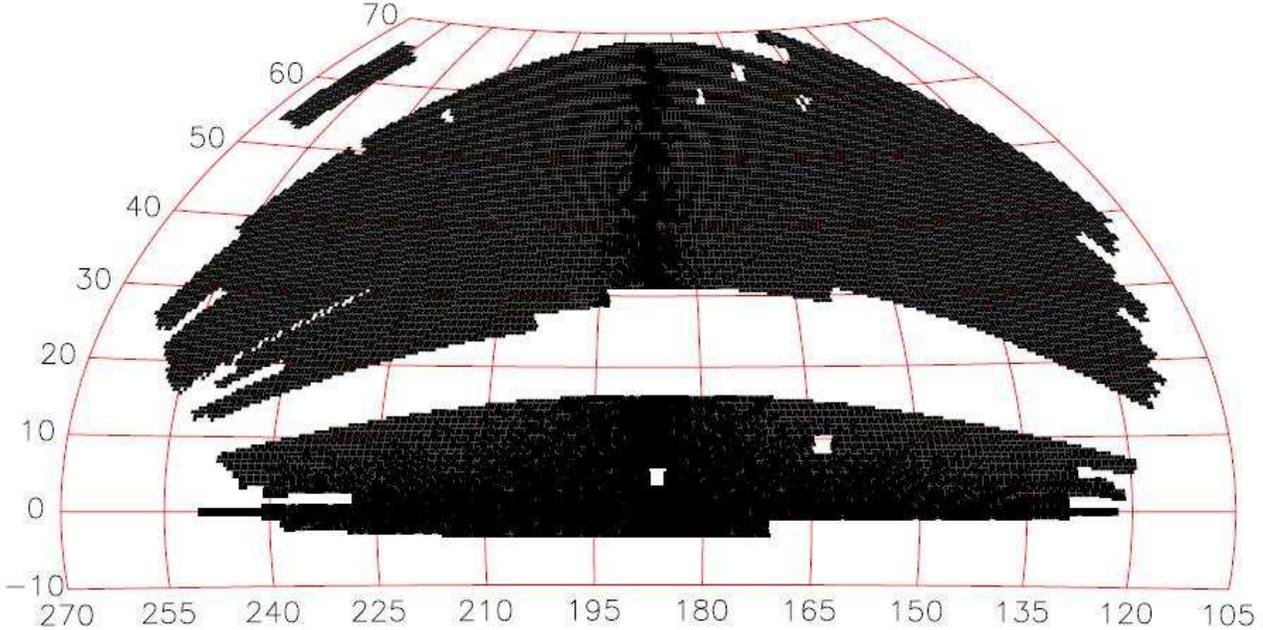,width=17cm,angle=0}
\caption{The angular coverage map of the NGP region of the SDSS DR4,
  which was analyzed in this study.  The right ascension and
  declination values are plotted in degrees and an equal-area Aitoff
  projection is used.  The total area of this survey region is
  $5{,}914$ deg$^2$.  The dashed lines indicate constant Galactic
  latitudes of $45^\circ$ (the innermost pair), $30^\circ$ and
  $15^\circ$.  There are several small gaps within the observed
  regions due to faults or poor-quality data.}
\label{figwindow}
\end{figure*}

We applied some star-galaxy separation criteria to the resulting
catalogue (Collister et al.\ 2006).  Firstly we made cuts based on
galaxy angular size and the difference between the ``p.s.f.'' and
model magnitudes.  In addition, we trained our photometric-redshift
neural network to improve the efficiency of star-galaxy separation.
We introduced an additional parameter $\delta_{\rm sg}$ to the network
such that $\delta_{\rm sg} = 1$ if the training 2SLAQ object is a
galaxy and $\delta_{\rm sg} = 0$ if the object has been
spectroscopically identified as a star.  The network then output
values of $\delta_{\rm sg}$ for each photometric object.  These values
form a continuous distribution between 0 and 1 and may be interpreted
as a star-galaxy classification probability.  Catalogue galaxies with
low values of $\delta_{\rm sg}$ were then excluded from the clustering
analysis.  We chose the cut $\delta_{\rm sg} > 0.2$, which reduced the
level of stellar contamination to $\sim 1.5\%$ with the loss of only
$\sim 0.1\%$ of the genuine galaxies.  The final catalogue analyzed in
this paper contains $644{,}903$ entries.  In Section \ref{secsys} we
consider the effects on the power spectrum of varying the star-galaxy
separation procedure.

\subsection{Angular selection function}

We measured the clustering of galaxies selected from the SDSS 4th Data
Release (DR4) in the North Galactic Plane (NGP).  We analyzed galaxies
in the area bounded by right ascensions $110^\circ < \alpha <
270^\circ$ and declinations $-5^\circ < \delta < 70^\circ$.  This area
excludes the three SDSS stripes in the South Galactic Plane (stripes
76, 82 and 86) which contribute a small fraction of the surveyed area
and which are widely spaced from the rest of the survey region.

We determined the angular selection function within the region
$(110^\circ < \alpha < 270^\circ, -5^\circ < \delta < 70^\circ)$ using
the coverage mask provided by the SDSS team (the files {\tt
  atStripeDef.par} and {\tt tsChunk.dr4.best.par} available via the
web-page {\tt http://www.sdss.org/dr4/coverage}).  The
publicly-available information includes the extent of the observations
in each survey stripe and a list of holes due to faults or
poor-quality data.  This information is given in terms of SDSS
``survey co-ordinates'' ($\mu$ and $\nu$ for each stripe) which may be
converted to right ascension and declination using the relations given
on the web-page.  This data was used to produce our initial window
function by assigning ``1'' to the surveyed areas and ``0'' to the
unsurveyed areas (Figure \ref{figwindow}).  The sky area encompassed
by this window function is $5{,}914$ deg$^2$, mapping out a total
cosmic volume $2.5 \, h^{-3}$ Gpc$^3$ across the redshift range $0.4 <
z < 0.7$.

There are various effects which may cause the true angular selection
function for the survey data to deviate from this binary coverage map
(Scranton et al.\ 2005).  These effects include: varying completeness
in the overlap regions between survey stripes, dust extinction or
seeing variations, incompleteness in the vicinity of very bright stars
or galaxies, and systematic effects induced by star-galaxy separation.
If left uncorrected, these effects may potentially cause a systematic
error in the measured clustering pattern, particularly on large
scales.  We investigate the importance of these effects in Section
\ref{secsys} using an extensive series of tests.  {\it Our conclusion
  is that, as far as we can tell from these tests, the binary coverage
  map is an acceptable approximation in the galaxy magnitude range
  $17.5 \le i_{\rm deV} \le 19.8$ analyzed in this study.}

\subsection{Photometric redshifts}
\label{secannz}

We determined photometric redshift estimates for the LRG catalogue
using the software package ``{\tt ANNz}'' (Collister \& Lahav 2004).
In brief, artificial neural networks are applied to parameterize the
relation between the galaxy redshift and the input information
(principally the galaxy photometry, but other data such as the angular
radius containing a given fraction of the galaxy flux can be readily
included).  The neural network is ``trained'' using a set of galaxies
with known spectroscopic redshifts (in this case, a subset of the
2SLAQ database) by minimizing a ``cost function'' (essentially, the
sum of the squared differences between the photometric and
spectroscopic redshifts).  After each training iteration, the cost
function is also evaluated for a separate ``validation'' set of
spectroscopic redshifts (a second subset of the 2SLAQ database) in
order to avoid over-fitting to the training set given the presence of
noise.  The final photometric-redshift performance is objectively
assessed by a third subset of the spectroscopic sample called the
``evaluation'' set.  Further details on the {\tt ANNz} software may be
found in Collister \& Lahav (2004).

For the input photometry we used the de-reddened $griz$-band ``model
magnitudes'' of the SDSS galaxies, which provide the most robust
colour estimates.  We excluded the $u$-band photometry owing to
concerns over a time-varying red leak in this filter that could
potentially introduce systematic errors in the photometric redshifts
as a function of galaxy position on the sky.  In fact, irrespective of
this concern, the $u$-band does not contribute any measurable benefit
to the photometric redshift accuracy owing to the relatively low
signal-to-noise ratio of the $u$-band photometry.

\begin{figure}
\center
\epsfig{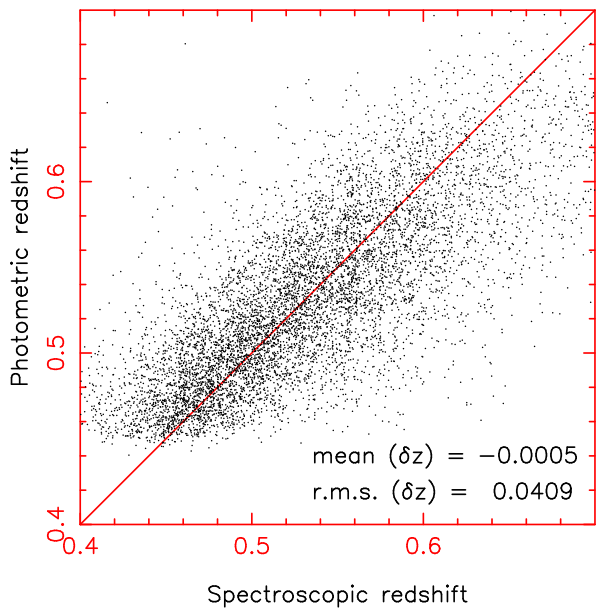}
\caption{Comparison between the photometric redshift $z_{\rm phot}$
  and spectroscopic redshift $z_{\rm spec}$ of galaxies in our
  evaluation set (for the conservative version of the catalogue
  analyzed in this paper).  The mean and standard deviation of the
  quantity $\delta z = z_{\rm phot} - z_{\rm spec}$ are indicated.}
\label{figzphotspec}
\end{figure}

The resulting photometric redshifts for the conservative version of
the catalogue analyzed in this paper, assessed using the spectroscopic
evaluation set, possess a standard redshift error
\begin{equation}
\sigma_z \equiv \left[ \overline{ (\delta z)^2 } - \left( \overline{
    \delta z } \right)^2 \right]^{1/2} \approx 0.041
\end{equation}
where $\delta z = z_{\rm phot} - z_{\rm spec}$, and do not suffer from
any significant bias in the mean difference with respect to the
spectroscopic redshifts, $\overline{\delta z} < 0.001$ (considering
galaxies in the range $0.4 < z_{\rm spec} < 0.7$, $0.4 < z_{\rm phot}
< 0.7$ and $i_{\rm deV} < 19.8$).  The performance of our code is
hence comparable to the results obtained for a similar sample of
galaxies by Padmanabhan et al.\ (2005).  Figure \ref{figzphotspec} is
a scatter plot of photometric and spectroscopic redshifts in the range
of interest.  The principal systematic effect is that galaxies with
$z_{\rm spec} < 0.45$ are scattered up in photometric redshift such
that there are very few galaxies with $z_{\rm phot} < 0.45$.  The
robustness of the photometric redshifts worsens in the vicinity of $z
\approx 0.4$ because the 4000\AA\, break enters a region of poor
sensitivity between the $g$ and the $r$ bands.  In addition, galaxies
with $z_{\rm spec} > 0.65$ are typically scattered down in redshift.
Photometric redshift performance improves steadily with brightening
magnitudes (and hence decreasing redshifts), reaching $\sigma_z
\approx 0.02$ for the brightest galaxies.  We refer the reader to
Collister et al.\ (2006) for more discussion.  We note that the small
level of systematic photo-$z$ bias evident in Figure
\ref{figzphotspec} is not a concern because it can be quantified using
the spectroscopic redshift data and folded into our survey
simulations.

\begin{figure}
\center
\epsfig{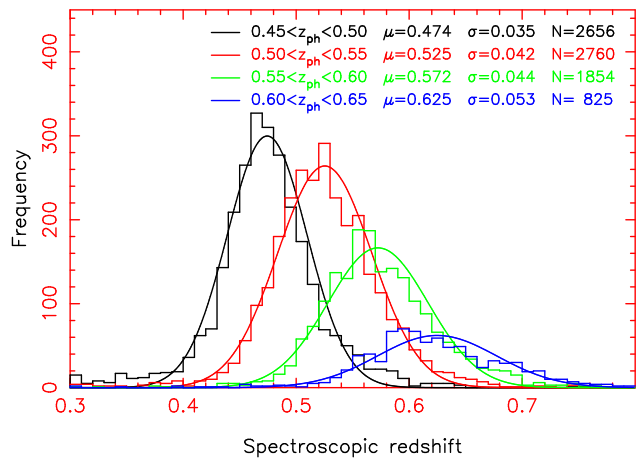}
\caption{Gaussian fits to the spectroscopic redshift distribution of
  Luminous Red Galaxies in each of four photometric redshift slices
  analyzed in this study.  The best-fitting Gaussian parameters $\mu$
  and $\sigma$ are indicated, where $p(z) \propto
  \exp{\{-[(z-\mu)^2/2\sigma^2]\}}$.  $N$ is the number of galaxies in
  the evaluation set used to measure the distribution in each slice.}
\label{fignzcal}
\end{figure}

Figure \ref{fignzcal} is a plot of the distribution of spectroscopic
redshifts in the evaluation set for galaxies binned in four
photometric-redshift slices of width $\Delta z = 0.05$ from $z = 0.45$
to $z = 0.65$.  We find that these error distributions are well fitted
by Gaussian distributions; the best-fitting values of the mean and
standard deviation for the different slices are indicated in the
Figure.  These Gaussian functions are taken as our model for the
redshift distribution of galaxies in each photo-$z$ slice when
analyzing the galaxy clustering results.  We repeated our cosmological
analyses using the raw binned redshift distributions in each slice in
place of the Gaussian fits, and found no significant difference in
results.  The number of galaxies with photometric redshifts outside
the range $0.45 < z < 0.65$ is very small.

The photometric redshift performance is sufficiently good that we can
observe the clustering visually.  Figure \ref{figradec} plots the 2000
most luminous galaxies in the stripe $135^\circ < \alpha < 235^\circ$,
$0^\circ < \delta < 10^\circ$ in each of a series of narrow redshift
slices of width $\Delta z = 0.03$.  The characteristic patterns of
large-scale structure may be readily discerned.

\section{Measurement of the angular power spectrum}
\label{secangpow}

\subsection{Method}
\label{secclest}

A distribution of galaxies can be related to its angular power
spectrum $C_\ell$ in two statistical steps.  Firstly, the galaxy
density field projected onto the sky, $\sigma(\theta,\phi)$, is
expanded in terms of its spherical harmonic coefficients $a_{\ell,m}$:

\begin{figure*}
\center
\epsfig{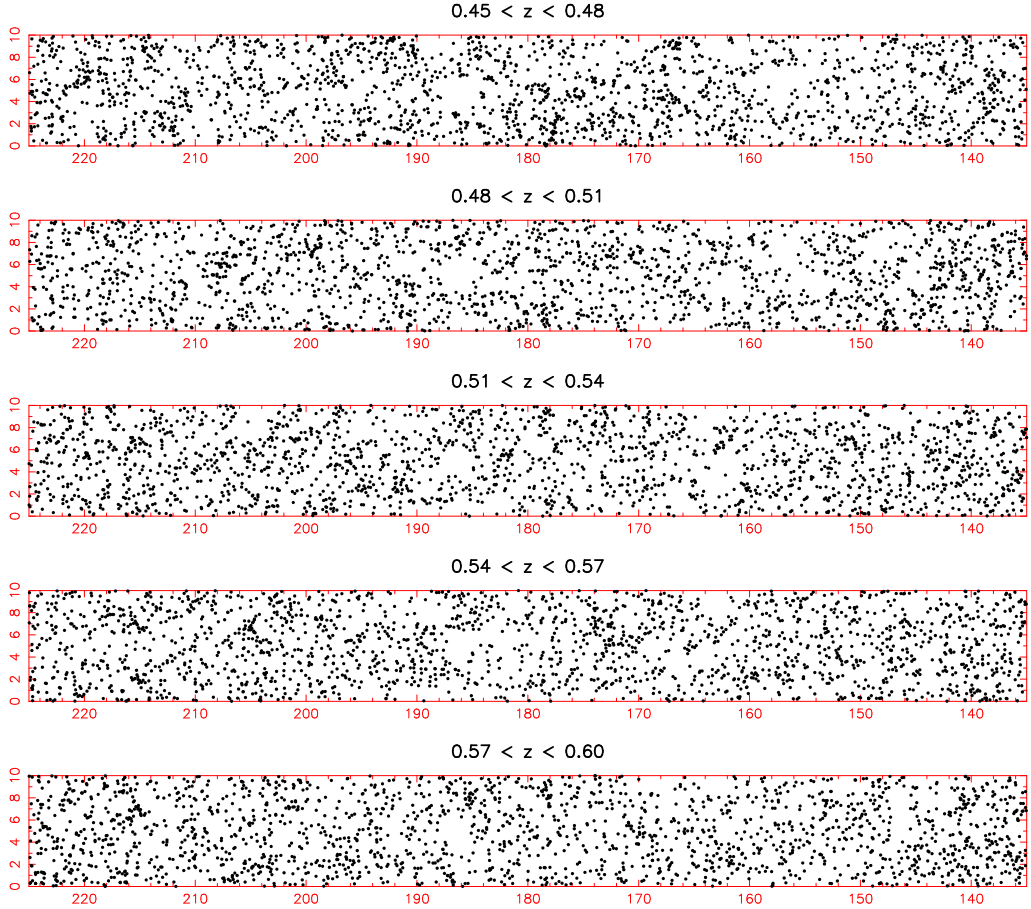}
\caption{The locations of the 2000 most luminous galaxies in each of a
  series of photometric redshift slices of width $\Delta z = 0.03$ in
  a narrow survey stripe.  The characteristic patterns of large-scale
  structure may be observed.  The axes are right ascension and
  declination in decimal degrees.}
\label{figradec}
\end{figure*}

\begin{equation}
\sigma(\theta,\phi) = \sum_{\ell=0}^\infty \sum_{m=-\ell}^\ell a_{\ell,m} \, Y_{\ell,m}(\theta,\phi)
\label{eqalmdef}
\end{equation}
where $Y_{\ell,m}$ are the usual spherical harmonic functions.
Secondly, galaxy positions are generated in a Poisson process as a
(possibly biased) realization of this density field.  The angular
power spectrum $C_\ell$ prescribes the expectation values of the
spherical harmonic coefficients in the first step of this model.  It
is defined over many realizations of the density field (indicated by
angled brackets) as:
\begin{equation}
<|a_{\ell,m}|^2> = C_\ell \, .
\end{equation}

The measurement of the angular power spectrum from a set of discrete
galaxy positions has been discussed by many authors.  The available
techniques can be crudely divided into two principal approaches.
Firstly, the power spectrum can be derived by direct spherical
harmonic estimation, in which the harmonic coefficients are calculated
by summation over the galaxy angular positions (e.g.\ Peebles 1973;
Scharf et al.\ 1992; Wright et al.\ 1994; Wandelt, Hivon \& Gorski
2001).  Secondly, a maximum likelihood approach can be utilized
(similar to those commonly employed for the analysis of CMB
temperature and polarization maps).  Examples of the application of
maximum likelihood methods to galaxy clustering measurements are the
papers by Efstathiou \& Moody (2001); Huterer, Knox \& Nichol (2001)
and Tegmark et al.\ (2002).

We used the technique of spherical harmonic estimation in the present
study.  We give a brief summary of the method here, referring the
reader to Blake, Ferreira \& Borrill (2004) for more details.  The
spherical harmonic coefficients of a density field for a {\it
  completely observed sky} may be estimated by summing over the $N$
galaxy positions $(\theta_i,\phi_i)$:
\begin{equation}
A_{\ell,m} = \sum_{i=1}^N Y_{\ell,m}^*(\theta_i,\phi_i)
\label{eqalm}
\end{equation}
The estimator for the angular power spectrum is then
\begin{equation}
C_{\ell,m}^{\rm obs} = |A_{\ell,m}|^2 - N/\Delta \Omega
\end{equation}
such that $<C_{\ell,m}^{\rm obs}> = C_\ell$.  $\Delta \Omega$ is the
total survey area, which is $4\pi$ for a complete sky.  The
discreteness of the distribution causes the correction term
``$-N/\Delta\Omega$''.  For a given multipole $\ell$ there are
$2\ell+1$ different estimators of $C_\ell$ corresponding to
$m=-\ell,-\ell+1,...,-1,0,1,...,\ell$.

For an {\it incomplete sky}, these equations must be corrected for
the unsurveyed regions, such that an estimate of $C_\ell$ is
\begin{equation}
C_{\ell,m}^{\rm obs} = \frac{|A_{\ell,m} - (N/\Delta\Omega) \,
  I_{\ell,m}|^2}{J_{\ell,m}} - \frac{N}{\Delta \Omega}
\label{eqclest}
\end{equation}
(e.g.\ Peebles 1973 equation 50) where
\begin{equation}
I_{\ell,m} = \int_{\Delta \Omega} Y_{\ell,m}^* \, d\Omega
\end{equation}
\begin{equation}
J_{\ell,m} = \int_{\Delta \Omega} |Y_{\ell,m}|^2 \, d\Omega
\end{equation}
where the integrals are performed over the survey area $\Delta
\Omega$, and are determined in our analysis by numerical integration.
For the SDSS area analyzed in this study (Figure \ref{figwindow}),
$\Delta \Omega / 4 \pi = 0.143$.

\begin{figure}
\center
\epsfig{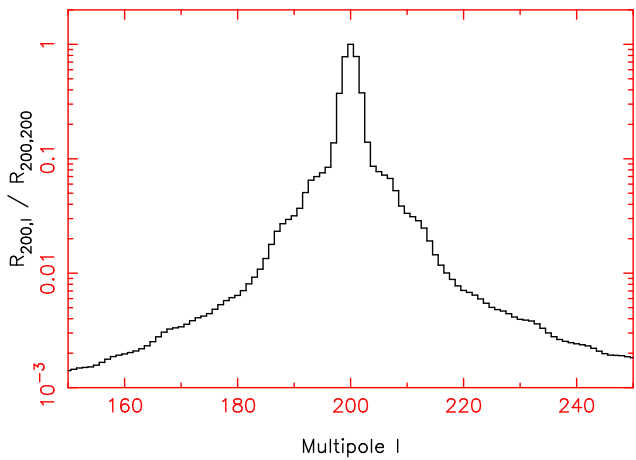}
\caption{The mixing matrix $R_{\ell,\ell'}$ for multipole $\ell = 200$
  as a function of $\ell'$, indicating the range of multipoles
  correlated by the survey window function effects (see equations
  \ref{eqclmix} and \ref{eqrll} for the definition of
  $R_{\ell,\ell'}$).}
\label{figclmix}
\end{figure}

We determined the observed angular power spectrum for the $\ell$th
multipole, $C_\ell^{\rm obs}$, by averaging equation \ref{eqclest}
over $m$:
\begin{equation}
C_\ell^{\rm obs} = \frac{\sum_{m=-\ell}^\ell C_{\ell,m}^{\rm
    obs}}{2\ell+1}
\label{eqclave}
\end{equation}
The partial sky coverage has the effect of convolving the harmonic
coefficients such that the measured angular power spectrum at
multipole $\ell$ depends on a range of multipoles $\ell'$ of the
underlying power spectrum:
\begin{equation}
< C_\ell^{\rm obs} > = \sum_{\ell'} R_{\ell,\ell'} \, C_{\ell'}
\label{eqclmix}
\end{equation}
The ``mixing matrix'' $R_{\ell,\ell'}$ may be determined from the
angular power spectrum $W_\ell$ of the survey window function:
\begin{equation}
R_{\ell,\ell'} = \frac{2 \ell' + 1}{4\pi} \sum_{\ell''} (2\ell'' + 1)
\, W_{\ell''} \left( \begin{array}{ccc} \ell & \ell' & \ell'' \\ 0 & 0
  & 0 \end{array} \right)^2
\label{eqrll}
\end{equation}
where the matrix in equation \ref{eqrll} is a Wigner 3-$j$ symbol and
\begin{equation}
W_\ell = \frac{\sum_{m=-\ell}^\ell |I_{\ell,m}|^2}{2\ell + 1}
\end{equation}
(e.g.\ Hivon et al.\ 2002; Deligny et al.\ 2004).  In Figure
\ref{figclmix} we plot normalized values of the mixing matrix
$R_{\ell,\ell'}$ evaluated using equation \ref{eqrll}, corresponding
to the SDSS DR4 geometry displayed in Figure \ref{figwindow}, for
multipole $\ell = 200$ as a function of $\ell'$.  The convolution
function $R_{200,\ell'}$ is sharply peaked, dropping to $10\%$ of its
peak value for an offset $\Delta \ell = 4$ from the peak.  The shape
of this function is largely insensitive to the choice of $\ell = 200$:
to a very good approximation, changing this value of $\ell$ only
causes a translation of the function along the $\ell'$-axis.

We averaged the measured angular power spectra in multipole bands of
width $\Delta \ell = 10$.  It can be seen from the foregoing argument
that these bands are approximately statistically independent
(correlated at a level below $10\%$) and we neglected the residual
correlations.  We determined the angular power spectrum up to a
maximum multipole $\ell_{\rm max} = 500$ (in Section \ref{secclfit} we
limit the range of multipoles to be fitted by our theoretical model in
accordance with the approximate extent of the linear clustering
regime).

\subsection{Error determination}

We compared various different determinations of the statistical error
$\sigma(C_\ell)$ in our measurement of the angular power spectrum.
Firstly, the application of Gaussian statistics leads to a simple
analytical approximation (e.g.\ Dodelson 2003 p.342 eq.11.27):
\begin{equation}
\sigma(C_\ell) = \sqrt{\frac{2}{f_{\rm sky}(2\ell+1)}} \left( C_\ell +
\frac{1}{N/\Delta\Omega} \right)
\label{eqclerrgauss}
\end{equation}
where $N/\Delta\Omega$ is the average source density (in units of
sr$^{-1}$) and $f_{\rm sky}$ denotes the fraction of sky covered by
the survey.  The structure of this equation can be readily understood
in crude terms.  The contribution to the error for each value of $m$
consists of a cosmic variance term ($C_\ell$) and a shot noise term
[$(N/\Delta\Omega)^{-1}$].  The averaging over $m$ reduces the
variance by a factor $2\ell+1$.  The fraction of sky covered by the
survey enters as $\sigma(C_\ell) \propto 1/\sqrt{f_{\rm sky}}$.  (If
the observed sky area is halved and just one of these sub-areas is
analyzed, we would expect the standard deviation of the measurement to
increase by a factor $\sqrt{2}$, under the approximation that the two
sub-areas are independent.)

We used equation \ref{eqclerrgauss} in our cosmological parameter
fitting.  We note that the value of $C_\ell$ in the equation could
correspond to either the {\it measured} power or the {\it model} power
at the multipole in question.  In our treatment we used the model
angular power spectrum for each combination of parameters to re-derive
the statistical error and assess the goodness-of-fit of that set of
parameters.

Estimating the measurement error using equation \ref{eqclerrgauss}
involves at least two approximations:
\begin{itemize}
\item The effects of the survey window function are encapsulated in
  just one quantity $f_{\rm sky}$, thus equation \ref{eqclerrgauss} is
  not likely to be valid for multipoles $\ell$ corresponding to
  angular scales $\theta \sim 180^\circ/\ell$ which are similar to or
  larger than those defining the structure of the window function (for
  which edge effects, i.e.\ the distribution of the survey area, are
  likely to be important).
\item Gaussian statistics are assumed, thus equation
  \ref{eqclerrgauss} is not likely to be valid in the regime of
  non-linear clustering.
\end{itemize}
However, it may be hoped that equation \ref{eqclerrgauss} is a good
approximation for multipoles $\ell$ corresponding to angular scales
$\theta$ smaller than the typical structure of the window function
($\theta < 2.5^\circ$ in our case), but large enough that the
corresponding spatial scales lie in the (quasi-)linear regime where
Gaussian statistics should apply.  This turns out to be the case for
our analysis.

\begin{figure}
\center
\epsfig{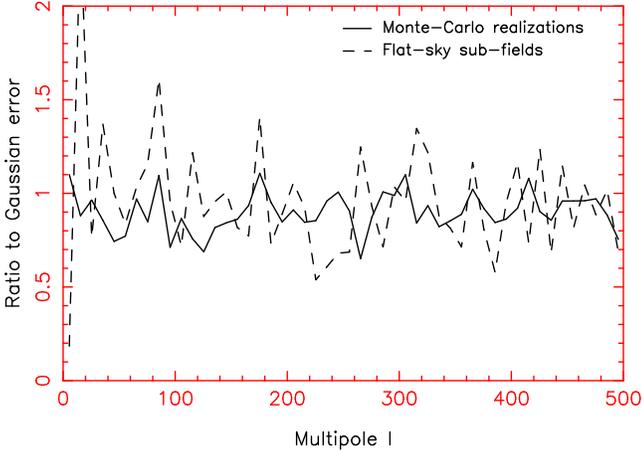}
\caption{The ratio of the error in the angular power spectrum
  measurement to the Gaussian prediction of equation
  \ref{eqclerrgauss} determined by two different techniques: firstly
  by Monte Carlo realizations (the solid line), and secondly by an
  analysis dividing the survey area into many independent patches (the
  dashed line).  The Gaussian prediction is a good approximation for
  the scales of interest.  The ``spikiness'' of the dashed line is
  caused by the poor signal-to-noise ratio of the measurements in the
  $10 \times 10$ deg patches, but the average behaviour is in good
  agreement with the Gaussian prediction.}
\label{figclerrcomp}
\end{figure}

In order to test the validity of equation \ref{eqclerrgauss}, we
estimated the error in the angular power spectrum using two additional
methods:
\begin{itemize}
\item We generated Monte Carlo realizations of galaxy distributions
  consistent with a model angular power spectrum similar to that
  measured for the real data, using the same survey window function as
  plotted in Figure \ref{figwindow}.  For each realization we
  calculated the density field $\sigma(\theta,\phi)$ (over a fine grid
  of pixels) by evaluating equation \ref{eqalmdef} for a set of
  spherical harmonic coefficients corresponding to a Gaussian
  realization of the underlying $C_\ell$ spectrum.  We then produced a
  catalogue of discrete galaxies as a Poisson realization of this
  density field, and measured the angular power spectrum using the
  same techniques utilized for the real survey data.  The scatter of
  the measured power spectra across these Monte Carlo realizations
  will include the full effects of the survey window function, unlike
  equation \ref{eqclerrgauss}, and this investigation is therefore
  designed to assess the approximation of using the single quantity
  $f_{\rm sky}$ to represent these effects.
\item Using a ``boot-strap'' approach, we divided the observed survey
  area into many $10 \times 10$ deg patches (the choice of patch size
  is a compromise between capturing a sufficient number of large-scale
  power spectrum modes within a single patch, and ensuring there are
  enough patches across the sky to permit reasonable statistics).  We
  measured the angular power spectrum of the galaxies within each
  patch using a flat-sky approximation (i.e., by taking a 2D Fast
  Fourier transform of the overdensity field $\delta_x$ and evaluating
  the 2D power spectrum $P_2(k) \equiv |\delta_k|^2$.  In the
  small-angle approximation, $C_\ell = P_2(\ell)$.)  The standard
  deviation of the power spectra across these patches is not sensitive
  to the global survey window function, but should trace non-Gaussian
  effects present in the data, thereby assessing the approximation of
  Gaussianity in equation \ref{eqclerrgauss}.
\end{itemize}
In Figure \ref{figclerrcomp} we plot the ratio of the angular power
spectrum error measured by these two techniques to the prediction of
equation \ref{eqclerrgauss}.  It can be seen that equation
\ref{eqclerrgauss} is a good approximation for the standard deviation
for the scales of interest, and it is used in our analysis.  Section
\ref{secalm} presents an additional test for the Gaussianity of the
clustering fluctuations.

\subsection{Results}

\subsubsection{Angular power spectra}

We measured the angular power spectrum of the LRGs (using the
estimator defined by equations \ref{eqclest} to \ref{eqclave}) in four
photometric redshift slices of width $\Delta z = 0.05$ between $z =
0.45$ and $z = 0.65$.  The results, averaged in multipole bands of
width $\Delta \ell = 10$, are listed in Table \ref{tabcl}, together
with the Gaussian errors of equation \ref{eqclerrgauss} (divided by
$\sqrt{\Delta \ell}$ to account for the binning).  For the purposes of
this Table the measured power spectra are used to assign the errors
with equation \ref{eqclerrgauss}.  When fitting cosmological models to
these power spectra in Section \ref{secclfit}, the model power spectra
are used to determine the Gaussian errors.  The surface densities
$N/\Delta\Omega$ of galaxies in the four slices are $(35.7, 29.1,
18.8, 8.7)$ deg$^{-2}$.  The results are also plotted in the left-hand
panels of Figure \ref{figclpanels}.

\subsubsection{Probability distribution of $A_{\ell,m}$}
\label{secalm}

The distribution of values of $A_{\ell,m}$ (the estimated spherical
harmonic coefficients defined by equation \ref{eqalm}) is an
interesting probe of the galaxy distribution (Peebles 1973; Hauser \&
Peebles 1973).  Consider first an unclustered distribution of points
with surface density $N/\Delta\Omega$ over a full sky.  The central
limit theorem ensures that the real and imaginary parts of $A_{\ell,m}
= \sum_i Y_{\ell,m}^*(i)$ are drawn independently from Gaussian
distributions such that the normalization satisfies $<|A_{\ell,m}|^2>
= N/\Delta\Omega$.  It is then easy to show that $x = |A_{\ell,m}|^2$
has an exponential probability distribution $P(x)$ for $m \ne 0$:
\begin{equation}
P(x) \, dx = \frac{\exp{(-x/\alpha)}}{\alpha} dx
\label{eqalmprob}
\end{equation}
where $\alpha = N/\Delta\Omega$.  For a partial sky, $|A_{\ell,m}|^2$
should be replaced by $x = |A_{\ell,m} - (N/\Delta\Omega)
I_{\ell,m}|^2/J_{\ell,m}$.  For a clustered distribution, $\alpha$
assumes the value $(N/\Delta\Omega) + C_\ell$ and equation
\ref{eqalmprob} holds {\it if each Fourier component has a
  randomly-assigned phase}.  The validity of equation \ref{eqalmprob}
is therefore a good cross-check of the applicability of the linear
regime, for which the random-phase assumption should apply (Chiang \&
Coles 2000).

\begin{figure*}
\center
\epsfig{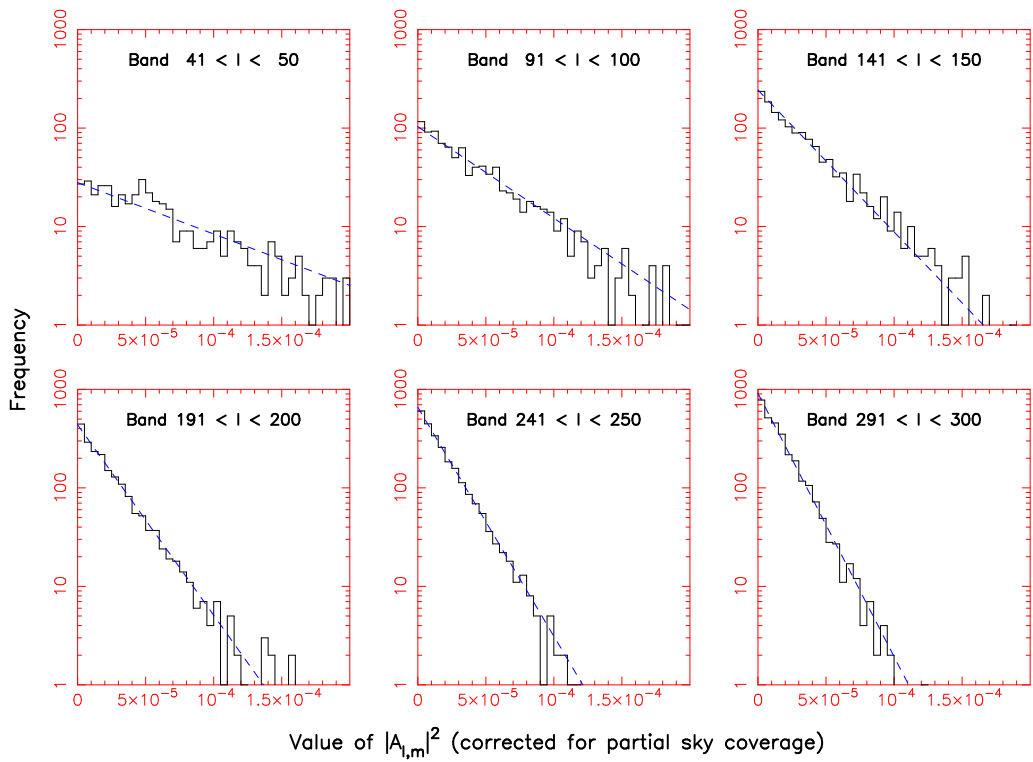}
\caption{The distribution of values of $|A_{\ell,m}|^2$ (corrected for
  partial sky coverage) for harmonic coefficients $1 \le m \le \ell$
  in six multipole bands of width $\Delta \ell = 10$ for the
  photometric redshift slice $0.45 < z < 0.5$.  The dashed lines are
  the predictions of random-phase statistics based on the observed
  strength of the power spectrum due to clustering and shot noise.
  The slope of this model changes across the six panels due to the
  varying amplitude of the angular power spectrum.}
\label{figalmhist}
\end{figure*}

Figure \ref{figalmhist} plots the distribution of the quantity $x$ for
harmonic coefficients in the range $1 \le m \le \ell$ in six different
multipole bands of width $\Delta \ell = 10$, for the first of the
photometric redshift slices defined above ($0.45 < z < 0.5$).  We also
display the expected probability distribution for random-phase
statistics, deriving the quantity $\alpha = (N/\Delta\Omega) + C_\ell$
from the galaxy surface density together with the measured strength of
the angular power spectrum in the multipole band.  We find that the
observed distribution of harmonic coefficients is a very good match to
the random-phase prediction even for the highest multipoles considered
($\ell = 300$).

\begin{table*}
\center
\caption{The angular power spectrum measurements in four photometric
  redshift slices of width $\Delta z = 0.05$ between $z = 0.45$ and $z
  = 0.65$.  This table lists the values of $10^5 \times C_\ell$ and
  the corresponding Gaussian error in multipole bands of width $\Delta
  \ell = 10$.  The errors $\sigma(C_\ell)$ are determined for the
  purposes of this Table by substituting the measured values of
  $C_\ell$ in equation \ref{eqclerrgauss} and {\it not} from the
  best-fitting power spectrum model.}
\begin{tabular}{ccccccccc}
\hline
$\ell$ & Slice 1 & & Slice 2 & & Slice 3 & & Slice 4 & \\
& $C_\ell$ & $\sigma(C_\ell)$ & $C_\ell$ & $\sigma(C_\ell)$ & $C_\ell$ & $\sigma(C_\ell)$ & $C_\ell$ & $\sigma(C_\ell)$ \\
\hline
  6 & 19.400 &  7.880 & 12.600 &  5.050 & 19.000 &  9.600 & 37.200 & 24.300 \\
 16 & 14.500 &  3.390 &  9.510 &  2.400 & 12.000 &  3.000 & 10.800 &  3.470 \\
 26 &  8.980 &  1.770 &  7.680 &  1.520 & 10.900 &  2.110 &  9.770 &  2.250 \\
 36 &  7.930 &  1.240 &  7.600 &  1.260 &  7.490 &  1.390 &  6.150 &  1.420 \\
 46 &  6.400 &  0.900 &  5.960 &  0.871 &  5.900 &  0.933 &  6.080 &  1.200 \\
 56 &  5.290 &  0.694 &  4.360 &  0.614 &  5.010 &  0.749 &  4.150 &  0.863 \\
 66 &  5.240 &  0.632 &  4.740 &  0.597 &  4.360 &  0.622 &  4.260 &  0.811 \\
 76 &  4.450 &  0.517 &  5.360 &  0.629 &  3.870 &  0.530 &  3.570 &  0.690 \\
 86 &  3.580 &  0.409 &  4.300 &  0.484 &  4.280 &  0.536 &  3.280 &  0.617 \\
 96 &  3.390 &  0.364 &  3.630 &  0.401 &  3.600 &  0.448 &  2.790 &  0.540 \\
106 &  3.370 &  0.345 &  3.250 &  0.350 &  3.370 &  0.408 &  2.740 &  0.510 \\
116 &  2.860 &  0.289 &  2.830 &  0.302 &  2.560 &  0.331 &  2.980 &  0.508 \\
126 &  2.350 &  0.239 &  2.290 &  0.250 &  2.020 &  0.272 &  2.050 &  0.417 \\
136 &  2.260 &  0.224 &  2.160 &  0.231 &  2.220 &  0.278 &  2.620 &  0.442 \\
146 &  2.300 &  0.219 &  1.880 &  0.202 &  2.070 &  0.259 &  1.580 &  0.357 \\
156 &  2.130 &  0.200 &  1.890 &  0.197 &  1.760 &  0.227 &  1.510 &  0.339 \\
166 &  1.880 &  0.178 &  1.420 &  0.160 &  1.640 &  0.212 &  1.750 &  0.342 \\
176 &  1.580 &  0.153 &  1.770 &  0.178 &  1.580 &  0.203 &  1.600 &  0.323 \\
186 &  1.730 &  0.159 &  1.620 &  0.164 &  1.340 &  0.182 &  1.180 &  0.289 \\
196 &  1.320 &  0.131 &  1.200 &  0.134 &  1.570 &  0.191 &  1.390 &  0.294 \\
206 &  1.290 &  0.125 &  1.190 &  0.130 &  1.200 &  0.165 &  1.320 &  0.282 \\
216 &  1.300 &  0.122 &  1.200 &  0.128 &  1.290 &  0.166 &  1.290 &  0.275 \\
226 &  1.160 &  0.112 &  1.180 &  0.124 &  1.240 &  0.159 &  1.080 &  0.256 \\
236 &  1.180 &  0.111 &  1.120 &  0.118 &  1.130 &  0.150 &  0.775 &  0.234 \\
246 &  0.948 &  0.096 &  0.841 &  0.101 &  0.853 &  0.132 &  1.070 &  0.245 \\
256 &  1.050 &  0.100 &  0.909 &  0.102 &  0.925 &  0.133 &  0.890 &  0.230 \\
266 &  0.972 &  0.094 &  0.915 &  0.101 &  0.933 &  0.131 &  0.592 &  0.212 \\
276 &  0.815 &  0.084 &  0.949 &  0.101 &  0.899 &  0.127 &  0.796 &  0.218 \\
286 &  0.702 &  0.077 &  0.821 &  0.092 &  1.030 &  0.131 &  1.130 &  0.230 \\
296 &  0.759 &  0.079 &  0.629 &  0.081 &  0.712 &  0.113 &  0.893 &  0.215 \\
306 &  0.809 &  0.079 &  0.776 &  0.087 &  0.578 &  0.105 &  0.535 &  0.194 \\
316 &  0.819 &  0.079 &  0.783 &  0.086 &  0.612 &  0.105 &  0.762 &  0.202 \\
326 &  0.794 &  0.076 &  0.715 &  0.082 &  0.679 &  0.107 &  0.646 &  0.193 \\
336 &  0.667 &  0.069 &  0.694 &  0.079 &  0.761 &  0.109 &  0.722 &  0.194 \\
346 &  0.726 &  0.071 &  0.701 &  0.078 &  0.588 &  0.099 &  0.904 &  0.199 \\
356 &  0.649 &  0.067 &  0.677 &  0.076 &  0.692 &  0.103 &  0.841 &  0.193 \\
366 &  0.593 &  0.063 &  0.574 &  0.071 &  0.623 &  0.098 &  0.659 &  0.183 \\
376 &  0.645 &  0.065 &  0.722 &  0.076 &  0.498 &  0.092 &  0.673 &  0.181 \\
386 &  0.568 &  0.061 &  0.518 &  0.067 &  0.490 &  0.090 &  0.457 &  0.169 \\
396 &  0.595 &  0.061 &  0.600 &  0.069 &  0.625 &  0.094 &  0.635 &  0.175 \\
406 &  0.547 &  0.058 &  0.546 &  0.066 &  0.514 &  0.089 &  0.622 &  0.172 \\
416 &  0.392 &  0.051 &  0.499 &  0.063 &  0.522 &  0.088 &  0.487 &  0.164 \\
426 &  0.584 &  0.058 &  0.566 &  0.065 &  0.454 &  0.084 &  0.433 &  0.160 \\
436 &  0.489 &  0.054 &  0.529 &  0.063 &  0.623 &  0.090 &  0.509 &  0.161 \\
446 &  0.521 &  0.054 &  0.432 &  0.059 &  0.461 &  0.083 &  0.599 &  0.163 \\
456 &  0.444 &  0.051 &  0.349 &  0.054 &  0.393 &  0.079 &  0.661 &  0.164 \\
466 &  0.452 &  0.051 &  0.361 &  0.054 &  0.515 &  0.083 &  0.566 &  0.159 \\
476 &  0.414 &  0.048 &  0.454 &  0.058 &  0.514 &  0.082 &  0.416 &  0.151 \\
486 &  0.397 &  0.047 &  0.473 &  0.058 &  0.322 &  0.074 &  0.352 &  0.147 \\
496 &  0.512 &  0.051 &  0.413 &  0.055 &  0.408 &  0.076 &  0.512 &  0.152 \\
\hline
\end{tabular}
\label{tabcl}
\end{table*}

\subsection{Cosmological parameter fits}
\label{secclfit}

\subsubsection{Independent redshift slices}

The angular power spectrum $C_\ell$ is a projection of the spatial
power spectrum of fluctuations at different redshifts $z$, $P(k,z)$,
where $k$ is a co-moving wavenumber.  The equation for the projection
is:
\begin{equation}
C_\ell = \frac{2 \, b^2}{\pi} \int P_0(k) \, g_\ell(k)^2 \, dk
\label{eqclexact}
\end{equation}
where $P(k,z) = P_0(k) \, D(z)^2$, with $D(z)$ the linear growth
factor at redshift $z$.  We note that this decomposition of $P(k,z)$
is strictly only valid in linear theory, and its application at
smaller scales is an approximation.  We have assumed a
scale-independent bias factor $b$ at redshift $z$ for the galaxies
with respect to the underlying matter fluctuations.

The kernel $g_\ell(k)$ is given by (neglecting redshift-space
distortions):
\begin{equation}
g_\ell(k) = \int_0^\infty j_\ell(u) \, f(u/k) \, du \, .
\label{eqker}
\end{equation}
Here, $j_\ell(x)$ is the spherical Bessel function and $f(x)$ depends
on the radial distribution of the sources as
\begin{equation}
f[z(x)] = p(z) \, D(z) \left( \frac{dx}{dz} \right)^{-1}
\end{equation}
where $x(z)$ is the co-moving radial co-ordinate at redshift $z$, and
$p(z)$ is the redshift probability distribution of the sources,
normalized such that $\int p(z) \, dz = 1$ (see e.g.\ Huterer, Knox \&
Nichol 2001; Tegmark et al.\ 2002; Blake et al.\ 2004).  A good
approximation for these equations which is valid for moderately large
$\ell \ga 50$ is:
\begin{equation}
C_\ell = b^2 \int P(k=\ell/x,z) \, x(z)^{-2} \, p(z)^2 \, \left(
\frac{dx}{dz} \right)^{-1} dz \, .
\label{eqclapprox}
\end{equation}

The angular power spectrum will be modified by {\it redshift-space
  distortions} on large scales.  These distortions significantly
affect the amplitude of the projected power spectrum on large scales
$\ell \la 50$, owing to the relative narrowness of each redshift
slice.  The amplitude of the redshift-space distortions is controlled
by a parameter $\beta(z) \approx \Omega_{\rm m}(z)^{0.6}/b(z)$, where
the quantities on the right-hand side of the equation are evaluated at
the centre of each redshift slice of our analysis.  The effect is to
introduce an additional term to the kernel of equation \ref{eqker}
such that it becomes $g_\ell(k) + g^\beta_\ell(k)$ where:
\begin{equation}
g^\beta_\ell(k) = \frac{\beta}{k} \int_0^\infty j_\ell^\prime(u) \,
f^\prime(u/k) \, du
\label{eqkerbeta}
\end{equation}
(Fisher, Scharf \& Lahav 1994; Padmanabhan et al.\ 2006 eq.26).

We can use these expressions to fit cosmological parameters to the
angular power spectrum data in each redshift slice.  The cosmological
parameters determine both the power spectrum of fluctuations $P(k,z)$
and the co-moving distance $x(z)$ in the above equations.  For the
redshift distribution of the sources $p(z)$ we used the Gaussian
functions fitted to the training set data (see Figure \ref{fignzcal}).
We derived model spatial power spectra using the ``{\tt CAMB}''
software package (Lewis, Challinor \& Lasenby 2000) which is based on
CMBFAST (Seljak \& Zaldarriaga 1996), including corrections for
non-linear growth of structure using the fitting formulae of Smith et
al.\ (2003) (``{\tt halofit=1}'' in {\tt CAMB}).  We outputted power
spectra at $z = 0.55$ (the mean redshift under analysis) and scaled
these spectra to other redshifts using the linear growth factor
$D(z)^2$.

\begin{figure}
\center
\epsfig{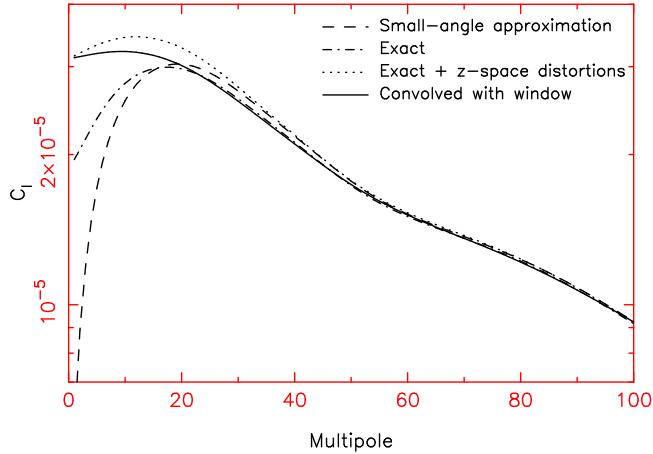}
\caption{The model angular power spectrum for the photometric redshift
  slice $0.45 < z < 0.5$ under various assumptions: (1) the
  small-angle approximation (equation \ref{eqclapprox}); (2) the exact
  expression in real space (equation \ref{eqclexact} with the kernel
  of equation \ref{eqker}); (3) after the inclusion of redshift-space
  distortions (with the addition of the kernel of equation
  \ref{eqkerbeta}); (4) after convolution with the survey window
  function using equation \ref{eqclmix}.  The assumed cosmological
  parameters are $\Omega_{\rm m} = 0.26$, $f_{\rm b} = 0.16$, $h =
  0.75$, $\sigma_8 = 1$, $\beta = 0.3$.  The small-angle approximation
  is acceptable for the multipole range $\ell \ga 50$; at low
  multipoles the results differ very significantly.}
\label{figclconv}
\end{figure}

\begin{figure}
\center
\epsfig{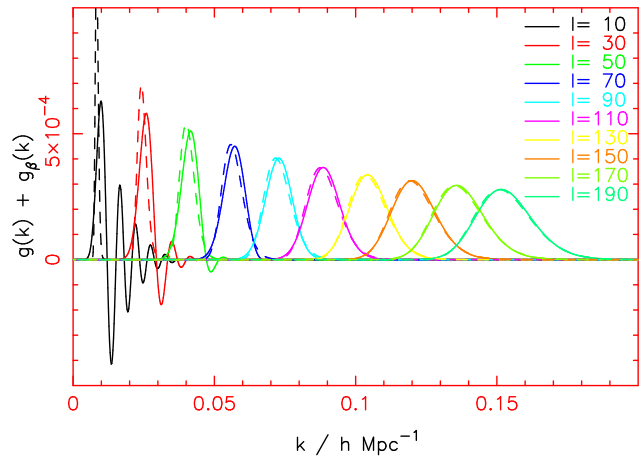}
\caption{The kernel $g_\ell(k) + g^\beta_\ell(k)$ appearing in
  equation \ref{eqclexact} for various multipoles $\ell$.  This plot
  indicates the range of Fourier scales $k$ contributing to angular
  power at a given multipole $\ell$ (each solid line is for a
  different value of $\ell$ increasing from left to right).  The
  dashed line indicates the kernel implied by the
  small-angle-approximation of equation \ref{eqclapprox}.  The assumed
  cosmological parameters are $\Omega_{\rm m} = 0.26$, $f_{\rm b} =
  0.16$, $h = 0.75$, $\sigma_8 = 1$, $\beta = 0.3$.  We again see that
  the small-angle approximation is acceptable for the multipole range
  $\ell \ga 50$.}
\label{figclker}
\end{figure}

For low multipoles $\ell$, we calculated the model angular power
spectrum for a given set of cosmological parameters using the exact
expression of equation \ref{eqclexact}, incorporating redshift-space
distortions.  For reasons of speed, we substituted the expression of
equation \ref{eqclapprox} at higher values of $\ell$ when the
approximation became acceptable.  Before comparing each model with
observations we convolved it with the survey window function using
equation \ref{eqclmix}.  Figure \ref{figclconv} displays the
modifications to the angular power spectrum caused by redshift-space
distortions and by the survey window function for an example case.
Figure \ref{figclker} plots the kernel $g_\ell(k)$ appearing in
equation \ref{eqclexact} (including the redshift-space distortion
contribution) for various multipoles $\ell$.  This kernel controls
which physical scales $k$ (in Fourier space) contribute power to a
clustering measurement at a particular multipole $\ell$.  It can be
seen that the projection at any given multipole involves a relatively
small range $\Delta k$ of scales owing to the narrowness of the
photometric-redshift slice $\Delta x$ in co-ordinate space (the
approximate range is $\Delta k/k = \Delta x/x$, where $k = \ell/x$).

The galaxy clustering pattern is well-known to be sensitive
principally to the quantity $\Omega_{\rm m} H_0$ (where $\Omega_{\rm
  m}$ is the matter density and $H_0$ is Hubble's constant), and
secondarily to the baryon fraction $\Omega_{\rm b}/\Omega_{\rm m}$.
For our initial investigation we fixed the values of the Hubble
constant $H_0 = 75$ km s$^{-1}$ Mpc$^{-1}$ and the tilt of the
primordial power spectrum $n_{\rm scalar} = 1$ (e.g.\ following Cole
et al.\ 2005; we consider variations in $H_0$ and $n_{\rm scalar}$ in
Section \ref{secclcomb}).  We then varied over a grid the matter
density $\Omega_{\rm m}$, the baryon fraction $f_{\rm b} = \Omega_{\rm
  b}/\Omega_{\rm m}$, the present-day normalization of the power
spectrum $\sigma_8$, and the constant galaxy bias factor $b$.  In the
linear regime $\sigma_8$ and $b$ are degenerate, but to allow for any
contribution from quasi-linear modes we marginalized over both
separately, using a flat prior $0.7 \le \sigma_8 \le 1.1$.  We
determined the co-moving distance $x(z)$ assuming a flat universe with
the remainder of the energy density provided by a cosmological
constant.  We calculated the value of the redshift-distortion
parameter at each model grid point by assuming $\beta = \Omega_{\rm
  m}(z_0)^{0.6}/b$, where $z_0$ is the central redshift of the
photo-$z$ slice.

We determined the relative likelihood of each model using:
\begin{equation}
{\rm Likelihood} \propto |M|^{-1/2} \exp{[-(d^T M^{-1} d)/2]}
\end{equation}
where $d = C_{\ell,{\rm obs}} - C_{\ell,{\rm mod}}$ is the vector of
differences between the model predictions and the data points in each
band of multipoles; and $M$ is the covariance matrix, evaluated for
the particular model at the grid point using equation
\ref{eqclerrgauss} (the chi-squared statistic is $\chi^2 = d^T M^{-1}
d$).  Since we are neglecting correlations between multipole bands,
the covariance matrix here is diagonal.

\begin{figure*}
\center
\epsfig{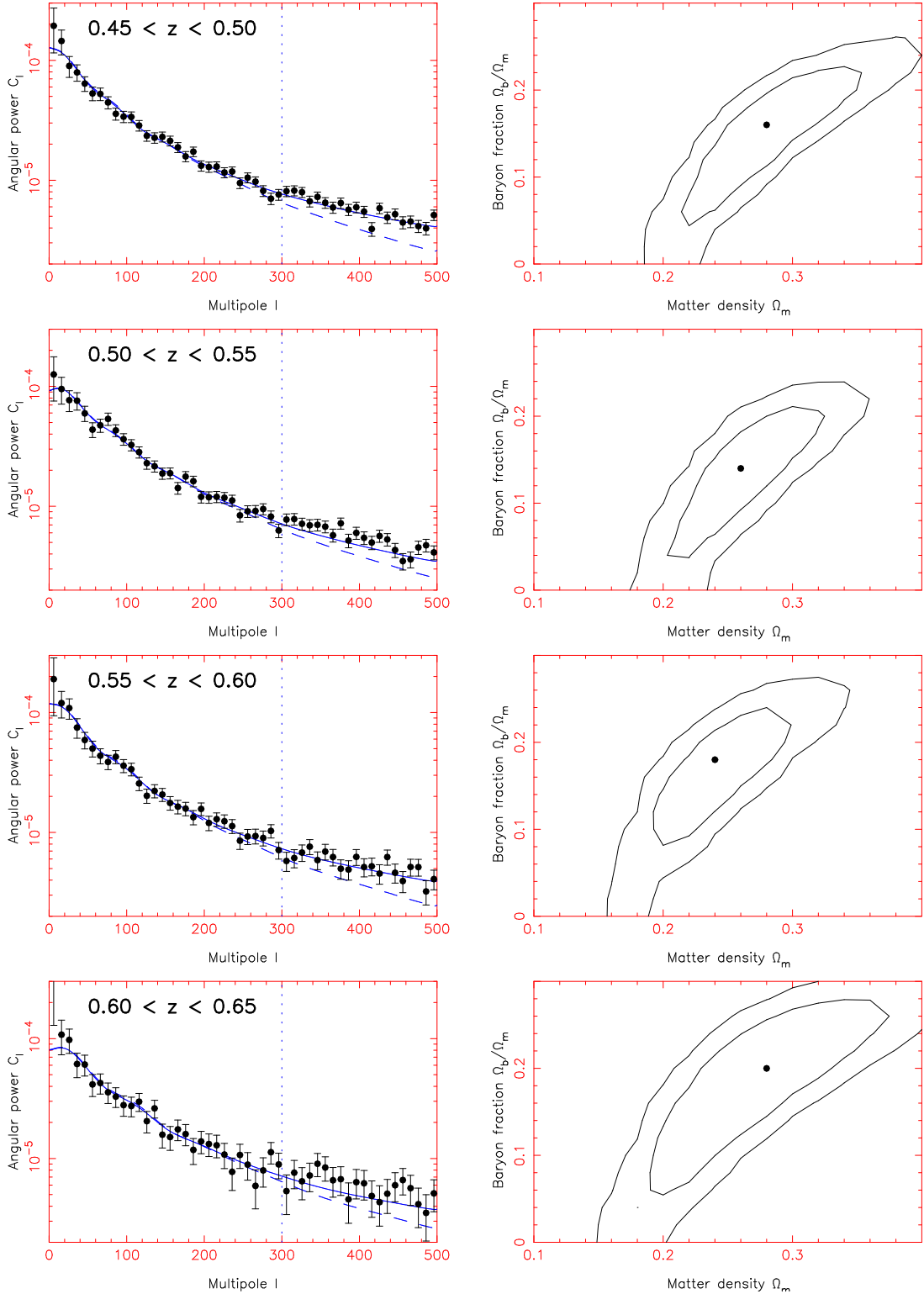}
\caption{Each row of this Figure corresponds to a different
  photometric redshift slice, indicated by the caption.  The left-hand
  column plots the angular power spectrum data (binned in multipole
  bands of width $\Delta \ell = 10$) and the best-fitting model in
  each case (solid line) including both redshift-space distortions and
  convolution with the survey window function.  The dashed line
  indicates a model with the same cosmological parameters but
  excluding the non-linear correction.  The vertical dotted line
  indicates the range of multipoles we used in our fitting, $\ell <
  \ell_{\rm max} = 300$.  This corresponds to an estimate of the
  extent of the linear regime: spatial scales $k \la 0.2 \, h$
  Mpc$^{-1}$.  The right-hand column displays 1-$\sigma$ and
  2-$\sigma$ probability contours in the plane of $\Omega_{\rm
    b}/\Omega_{\rm m}$ and $\Omega_{\rm m}$, marginalizing over
  $\sigma_8$ and $b$.  The solid circle indicates the best-fitting
  parameter combination in each case.  The cross displays the
  best-fitting result when the four redshift slices are combined with
  appropriate covariances.}
\label{figclpanels}
\end{figure*}

We assumed that power spectrum modes with spatial scales $k \la 0.2 \,
h$ Mpc$^{-1}$ could be modelled via equation \ref{eqclapprox} using a
scale-independent bias factor $b$ (this is comparable to the smallest
scales fitted in lower-redshift surveys -- $k_{\rm max} \approx 0.15
\, h$ Mpc$^{-1}$ in Percival et al.\ 2001 and $k_{\rm max} \approx 0.2
\, h$ Mpc$^{-1}$ in Tegmark et al.\ 2004a).  We estimated the
equivalent maximum multipole of the angular power spectrum using the
fact that at redshift $z$, a multipole $\ell$ probes spatial power on
scales of roughly $k = \ell/r(z)$ (see equation \ref{eqclapprox} and
Figure \ref{figclker}).  This corresponds to $\ell_{\rm max} \approx
300$ at our mean redshift $z \approx 0.55$ and hence for each redshift
slice we fitted our models to only the first 30 multipole bands of
width $\Delta \ell = 10$.  Our placement of the non-linear transition
at these angular scales is justified in Figure \ref{figclpanels} by
the small differences for $\ell < \ell_{\rm max}$ between the
best-fitting non-linear power spectrum in each slice (generated by
``{\tt halofit = 1}'' in {\tt CAMB}) and the corresponding linear
power spectrum with the same cosmological parameters (generated by
``{\tt halofit = 0}'' in {\tt CAMB}).

\begin{table*}
\center
\caption{Best-fitting values of the cosmological parameters for the
  modelling of the angular power spectra for multipoles $\ell \le
  \ell_{\rm max} = 300$ in the four photometric redshift slices.  The
  best-fitting values and errors for $\Omega_{\rm m}$ and $f_{\rm b} =
  \Omega_{\rm b}/\Omega_{\rm m}$ are obtained by marginalizing over
  the other three parameters, including $\sigma_8$ and $b$.  The
  values of the Hubble parameter $h$ and the tilt of the primordial
  power spectrum $n_{\rm scalar}$ are both held fixed in each analysis
  at the values given in the Table.  The best-fitting values of the
  linear bias $b$ are quoted assuming $\sigma_8 = 1$.  We also quote
  the value of the $\chi^2$ statistic for the best-fitting model and
  the number of degrees of freedom (d.o.f.) in each case, as well as
  the average galaxy absolute magnitude $M_i$.  The combined
  measurements of all four slices are obtained by calculating a
  covariance matrix between redshift slices as described in the text.}
\begin{tabular}{ccccccccc}
\hline
Slice & $h$ (fixed) & $n_{\rm scalar}$ (fixed) & $\Omega_{\rm m}$ &
$f_{\rm b}$ & $b$ & $M_i - 5$ log$_{10}h$ & $\chi^2$ & d.o.f. \\
\hline
$0.45 < z < 0.5$ & 0.75 & 1.00 & $0.28 \pm 0.05$ & $0.16 \pm 0.06$ & $1.51 \pm 0.03$ & $-22.45$ & 18.8 & 26 \\
$0.5 < z < 0.55$ & 0.75 & 1.00 & $0.26 \pm 0.04$ & $0.14 \pm 0.06$ & $1.68 \pm 0.05$ & $-22.67$ & 22.4 & 26 \\
$0.55 < z < 0.6$ & 0.75 & 1.00 & $0.24 \pm 0.04$ & $0.18 \pm 0.06$ & $1.74 \pm 0.06$ & $-22.92$ & 18.7 & 26 \\
$0.6 < z < 0.65$ & 0.75 & 1.00 & $0.28 \pm 0.06$ & $0.20 \pm 0.07$ & $1.95 \pm 0.12$ & $-23.17$ & 14.8 & 26 \\
\hline
Combined & 0.75 & 1.00 & $0.26 \pm 0.031$ & $0.16 \pm 0.036$ & & & 82.1 & 113 \\
& 0.7 & 1.00 & $0.28 \pm 0.032$ & $0.16 \pm 0.038$ & & & 82.3 & 113 \\
& 0.75 & 0.95 & $0.28 \pm 0.031$ & $0.16 \pm 0.039$ & & & 81.9 & 113 \\
\hline
\end{tabular}
\label{tabpar}
\end{table*}

The best-fitting values of $\Omega_{\rm m}$ and $f_{\rm b}$ at each
redshift, marginalizing over the other three parameters including
$\sigma_8$ and $b$, are listed in Table \ref{tabpar}.  The results --
$\Omega_{\rm m} \approx 0.27 \pm 0.05$ and $f_{\rm b} \approx 0.17 \pm
0.06$ -- are consistent across the redshift slices.  Our measurement
of $\sigma_8$ is weak because we are restricting ourselves mainly to
the linear clustering regime, where $\sigma_8$ and $b$ are largely
degenerate.  In Table \ref{tabpar} we quote the best-fitting values of
$b$ in each redshift slice assuming $\sigma_8 = 1$.  We find $1.5 < b
< 1.9$, with this high bias reflecting the fact that LRGs prefer
denser environments (e.g.\ Zehavi et al.\ 2005).  We note that (in the
linear regime) these bias measurements should be interpreted as
constraints on the product $\sigma_8 \times b$.

The bias systematically increases with redshift for two reasons:
\begin{enumerate}
\item Galaxies in more distant redshift slices are preferentially more
  luminous (owing to the fixed apparent magnitude threshold) and hence
  more strongly clustered (e.g.\ Norberg et al.\ 2002).  In Table
  \ref{tabpar} we list the average absolute $i$-band magnitude $M_i$
  of galaxies in each redshift slice, calculated using a K-correction
  obtained from a standard Luminous Red Galaxy template.
\item In standard models of the evolution of galaxy clustering, the
  bias factor of a class of galaxies increases with redshift in
  opposition to the decreasing linear growth factor, in order to
  reproduce the observed approximate constancy of the small-scale
  clustering length (e.g.\ Magliocchetti et al.\ 2000; Lahav et
  al.\ 2002).
\end{enumerate}

In Figure \ref{figclpanels} we plot the measured angular power spectra
in the four redshift slices together with the cosmological parameter
fits.  Each row corresponds to a different redshift slice as
indicated.  The left-hand column displays the data points and the
best-fitting models (solid line).  The dashed lines illustrate the
corresponding power spectrum models with the same cosmological
parameters but omitting the non-linear corrections.  The difference
between these curves indicates the importance of the non-linear
correction.  The vertical dotted line indicates the position of the
maximum multipole we used in our fitting, $\ell_{\rm max} = 300$.  The
right-hand column plots 1-$\sigma$ and 2-$\sigma$ probability contours
in the plane of $f_{\rm b}$ and $\Omega_{\rm m}$, marginalizing over
$\sigma_8$ and $b$.

In all redshift slices the best models produce good fits to the shape
of the clustering pattern over scales $\ell \le \ell_{\rm max} = 300$,
as indicated by the acceptable values of $\chi^2$ listed in Table
\ref{tabpar}.  In fact, the model generally remains a good match to
the data for smaller scales $\ell > 300$ which are not used in the
fitting.  At low multipoles there is a visual suggestion of a small
excess (1-$\sigma$) of measured power with respect to the model
prediction, but this statement lacks statistical significance.

\begin{figure}
\center
\epsfig{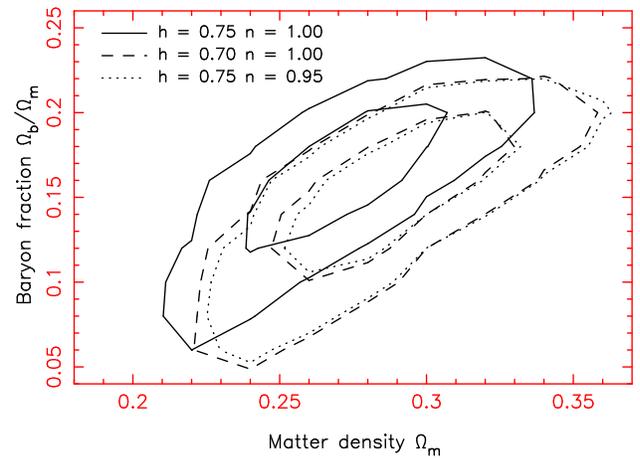}
\caption{The probability distribution for the cosmological parameters
  $\Omega_{\rm m}$ and $\Omega_{\rm b}/\Omega_{\rm m}$ combining the
  four redshift slices using a full covariance matrix, and
  marginalizing over $\sigma_8$ and the four linear bias parameters.
  1-$\sigma$ and 2-$\sigma$ contours are displayed.  The solid and
  dashed contours assume $H_0 = 75$ and $70$ km s$^{-1}$ Mpc$^{-1}$,
  respectively.}
\label{figcltotpars}
\end{figure}

\subsubsection{Combined redshift slices}
\label{secclcomb}

The cosmological results from different redshift slices are not
independent, because galaxies sampling the same cosmic variance are
scattered between slices by the photometric redshift errors.
Therefore we cannot combine the results from the different redshift
slices by simply multiplying together the four probability maps in
Figure \ref{figclpanels}.  Instead, we derived a full covariance
matrix between the measurements of the angular power spectrum in
multipole bands in all four redshift slices.

Since different multipole bands are uncorrelated within each redshift
slice (to a good approximation), the covariance matrix has a simple
structure of non-zero diagonals with entries for multipoles $\ell_i =
\ell_j$ for redshift slices $i$ and $j$.  These covariances are given
by
\begin{equation}
< C^i_\ell \, C^j_\ell > - < C^i_\ell > < C^j_\ell > = \frac{2}{f_{\rm
    sky} (2\ell + 1)} \, C^{i,j}_\ell
\end{equation}
using the same Gaussian approximation as equation \ref{eqclerrgauss}.
Here, $C^{i,j}_\ell$ is the cross angular power spectrum between the
two redshift slices.  This can be determined for a given cosmological
model by evaluating the projection equation \ref{eqclexact} including
a kernel $g_\ell(k)$ for each redshift slice:
\begin{equation}
C^{i,j}_\ell = \frac{2 \, b_i \, b_j}{\pi} \int P_0(k) \, g^i_\ell(k)
\, g^j_\ell(k) \, dk
\label{eqclcross}
\end{equation}
where $b_i$ and $b_j$ are the linear bias factors for the slices.

\begin{figure*}
\center
\epsfig{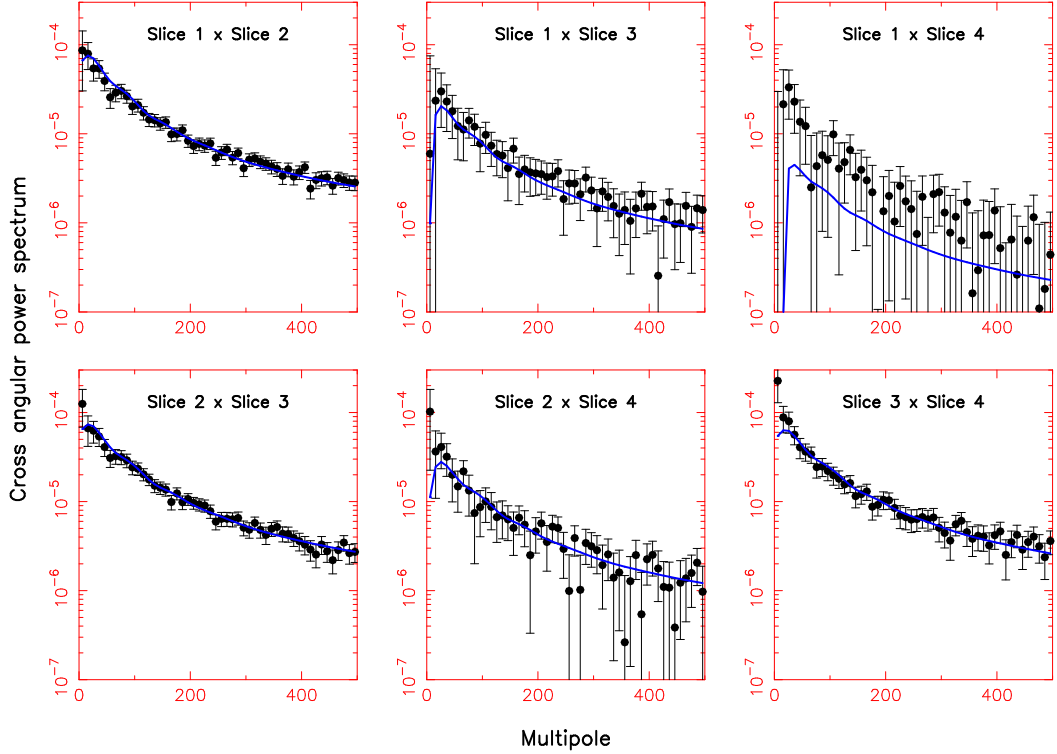}
\caption{The cross angular power spectra measured between all
  combinations of the four photometric redshift slices.  The plotted
  models are not fitted to the data points, but are calculated using
  equation \ref{eqclcross} assuming the best-fitting cosmological
  parameters from the analysis of the auto power spectra.}
\label{figclcross}
\end{figure*}

Using this form of covariance matrix, we re-fitted the cosmological
models as before.  We marginalized over a separate linear bias
parameter for each redshift slice ($b_1 \rightarrow b_4$), thus we are
fitting seven parameters ($\Omega_{\rm m}$, $f_{\rm b}$, $\sigma_8$,
$b_1 \rightarrow b_4$).  The probability contours for the matter and
baryon densities are displayed in Figure \ref{figcltotpars}.  The
resulting 1-$\sigma$ measurements of the parameters are $\Omega_{\rm
  m} = 0.26 \pm 0.031$ and $f_{\rm b} = 0.16 \pm 0.036$ (for fixed
values of $H_0 = 75$ km s$^{-1}$ Mpc$^{-1}$ and $n_{\rm scalar} = 1$,
marginalizing over $\sigma_8$ and the four bias parameters).  This
constitutes a significant detection of the influence of baryons on the
clustering pattern of galaxies.  The dashed contours in Figure
\ref{figcltotpars} illustrate the effect of reducing the value of
$H_0$ to 70 km s$^{-1}$ Mpc$^{-1}$.  The resulting shift is exactly as
expected given that the clustering pattern is sensitive to the
combination $\Omega_{\rm m} H_0$.  We therefore express our result as
$\Omega_{\rm m} h = 0.195 \pm 0.023$ [where $h = H_0/(100$ km s$^{-1}$
  Mpc$^{-1}$)].  The dotted contours in Figure \ref{figcltotpars} are
obtained if $H_0 = 75$ km s$^{-1}$ Mpc$^{-1}$ but the value of $n_{\rm
  scalar}$ is reduced to $0.95$.  In this case we obtain the
marginalized error $\Omega_{\rm m} h = 0.21 \pm 0.023$.

These parameter measurements agree with the latest analysis of the
Cosmic Microwave Background radiation (the 3-year results of the
Wilkinson Microwave Anisotropy Probe, see Spergel et al.\ 2006).  The
measurements from the CMB data alone are $\Omega_{\rm m} h \approx
0.17 \pm 0.03$ and $\Omega_{\rm b}/\Omega_{\rm m} \approx 0.18 \pm
0.01$.  Our results are also in concord with those derived from the
latest spectroscopic redshift surveys.  For example, Cole et
al.\ (2005) report $\Omega_{\rm m} h = 0.17 \pm 0.02$ and $\Omega_{\rm
  b}/\Omega_{\rm m} = 0.19 \pm 0.05$ from an analysis of the final 2dF
Galaxy Redshift Survey, fixing the Hubble parameter $h$ and scalar
index of primordial fluctuations $n_{\rm scalar}$ in a manner similar
to our analysis.  There is tension at the 1-$\sigma$ level between the
different best-fitting values for $\Omega_{\rm m}$, but we ascribe no
statistical significance to this.  Our parameter constraints are also
consistent with the combined analysis of the CMB data and power
spectrum of the spectroscopic sample of SDSS LRGs by Huetsi (2006b).

\subsubsection{Cross power spectra}

Equation \ref{eqclcross} can be used to evaluate the cross angular
power spectrum between different redshift slices $i$ and $j$ for a
given cosmological model.  The cross power spectrum can also be
measured from the data; this is a useful cross-check on the
consistency of our analysis (i.e.\ the best-fitting cosmological
parameters and the redshift distributions within each photo-$z$
slice).  The estimator is:
\begin{equation}
C^{i,j}_\ell = \frac{1}{2\ell + 1} \sum_{m=-\ell}^\ell
(A^i_{\ell,m})^* A^j_{\ell,m}
\end{equation}
where $A^i_{\ell,m}$ and $A^j_{\ell,m}$ are the spherical harmonic
coefficients estimated in redshift slices $i$ and $j$, corrected for
partial sky effects as described in Section \ref{secclest}.  By
analogy with equation \ref{eqclerrgauss}, the variance in this
estimator is:
\begin{equation}
\sigma^2(C^{i,j}_\ell) = \frac{2}{f_{\rm sky}(2\ell+1)} \left(
C^i_\ell + \frac{\Delta\Omega}{N_i} \right) \left( C^j_\ell +
\frac{\Delta\Omega}{N_j} \right)
\end{equation}
where $C^i_\ell$ and $C^j_\ell$ are the auto angular power spectra in
the two redshift slices and $N_i$ and $N_j$ are the numbers of
galaxies analyzed.

Figure \ref{figclcross} plots the measured cross power spectra for the
six possible combinations of four redshift slices, together with the
model predictions for the best-fitting cosmological parameters from
the combined analysis of the auto power spectra of all the redshift
slices.  Although the cross power spectra have not been used in the
cosmological fitting, the agreement is generally excellent.  The only
combination of redshift slices which performs poorly is 1 and 4, the
most separated pair.  In this case, the cross-correlation amplitude is
generated by the furthest wings of the redshift probability
distribution in each slice.  In this regime the Gaussian model falls
off too quickly, hence the model prediction lies below the cross power
spectrum data points.

\subsection{Searching for baryon oscillations}

\begin{figure}
\center
\epsfig{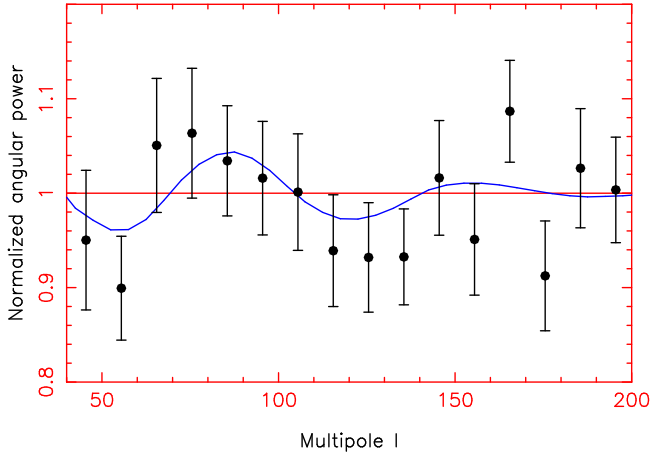}
\caption{Combination of the angular power spectrum measurements.  The
  data is first divided by a smooth polynomial fit and then scaled
  along the $x$-axis in order to match equivalent spatial scales
  projected from different redshifts.  The aim is to detect
  ``model-independent'' evidence for baryon oscillations in the
  clustering pattern.  The solid sinusoidal line is a model of the
  result.}
\label{figclsmooth}
\end{figure}

We used the angular power spectrum measurements to look for
``model-independent'' evidence of baryon oscillations in the
clustering pattern.  We divided out the overall shape of the power
spectrum in each case via a smooth polynomial fit, and searched for
significant enhancements or reductions in power at the expected
scales.  The baryon oscillations encode a preferred spatial scale
(fixed in co-moving co-ordinates) as an approximately sinusoidal
modulation of power in Fourier space.  Therefore, given data with a
sufficiently high signal-to-noise ratio, we should observe the
corresponding preferred angular scales moving to larger values of
$\ell$ with increasing redshift, which would constitute an excellent
``model-independent'' test for the existence of baryon oscillations.
Unfortunately, the current data does not possess a sufficiently high
signal-to-noise ratio to permit this test.  We therefore combined the
angular power spectrum measurements in the different redshift slices,
scaling a measurement at redshift $z$ along the $x$-axis by a factor
proportional to $1/x(z)$ in order to match equivalent spatial scales
projected from different redshifts.  One binning of the data is
plotted in Figure \ref{figclsmooth}, together with a model prediction.
We find visual suggestions of baryon oscillations, but the statistical
significance is less than 3-$\sigma$.  We have not taken into account
covariances between redshift slices when constructing Figure
\ref{figclsmooth}.

\section{Tests for systematic photometric errors}
\label{secsys}

In this Section we perform an extensive series of tests for potential
systematic photometric errors that may affect our clustering results.
Unfortunately, any small photometric errors will be amplified for a
sample of Luminous Red Galaxies because these objects are selected
from a very steep part of the galaxy luminosity function, where a
relatively small shift in the magnitude completeness threshold (e.g.,
$\Delta m = 0.05$) will produce a much larger (10s of per cent) change
in the galaxy surface density.  In order to search for such effects,
we compared the angular power spectrum measured for the ``default''
sample (defined in Section \ref{secdata}) with that obtained by
restricting or extending the galaxy selection in the following ways:

\begin{itemize}

\item Exclusion of areas of high dust extinction.

\item Exclusion of areas of poor astronomical seeing.

\item Exclusion of areas in the vicinity of very bright objects.

\item Exclusion of areas lying in the overlap regions between survey
  stripes.

\item Exclusion of areas of low Galactic latitude.

\item Separate analysis of the two largest disconnected survey regions
  in Figure \ref{figwindow}.

\item Variations in the star-galaxy separation criteria.

\end{itemize}

\begin{figure*}
\center
\epsfig{file=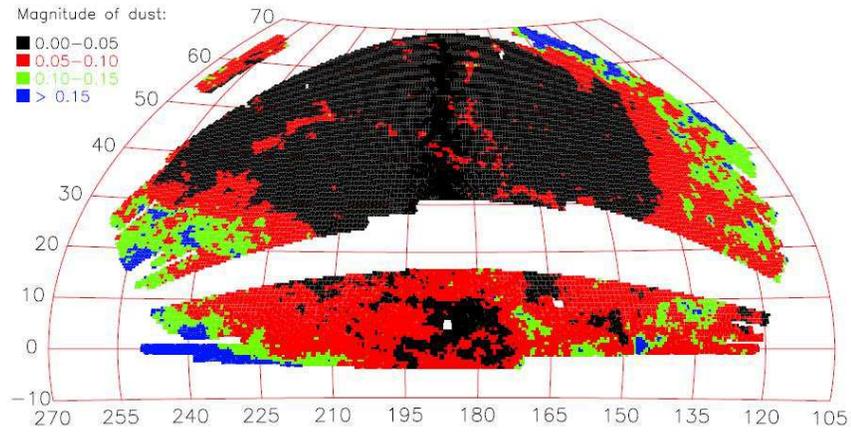,width=12cm,angle=0}
\caption{Variation in dust extinction across the DR4 NGP region.}
\label{figdust}
\end{figure*}

\begin{figure*}
\center
\epsfig{file=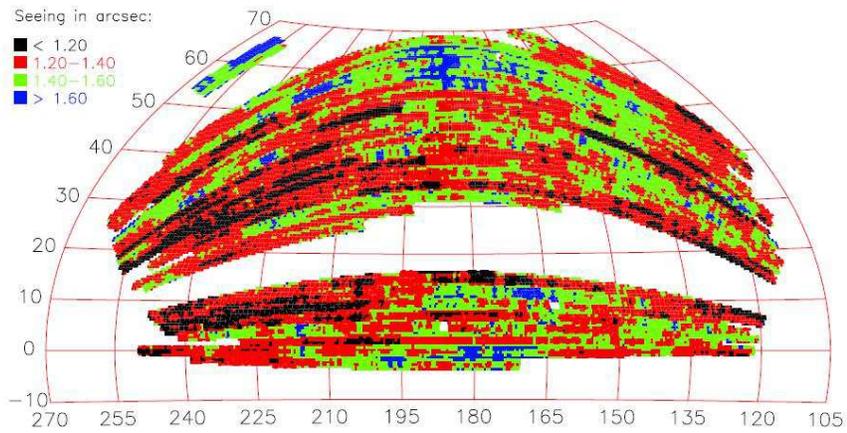,width=12cm,angle=0}
\caption{Variation in astronomical seeing across the DR4 NGP region.}
\label{figseeing}
\end{figure*}

\begin{figure*}
\center
\epsfig{file=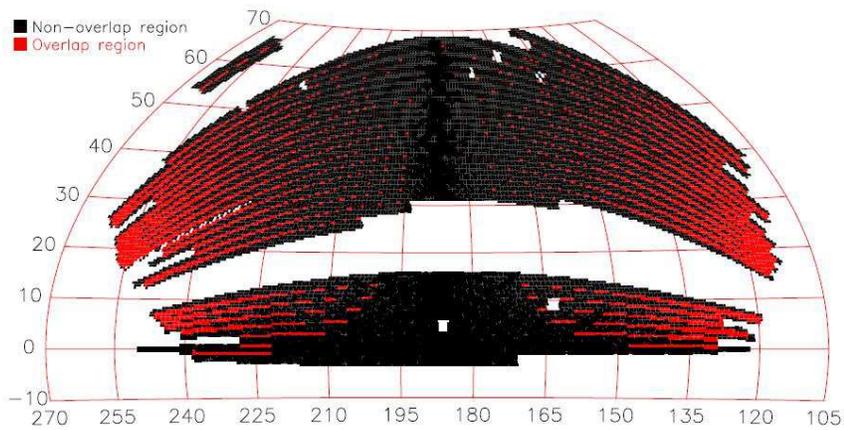,width=12cm,angle=0}
\caption{Location of overlap regions between SDSS stripes in the DR4
  NGP region.}
\label{figoverlap}
\end{figure*}

\begin{figure*}
\center
\epsfig{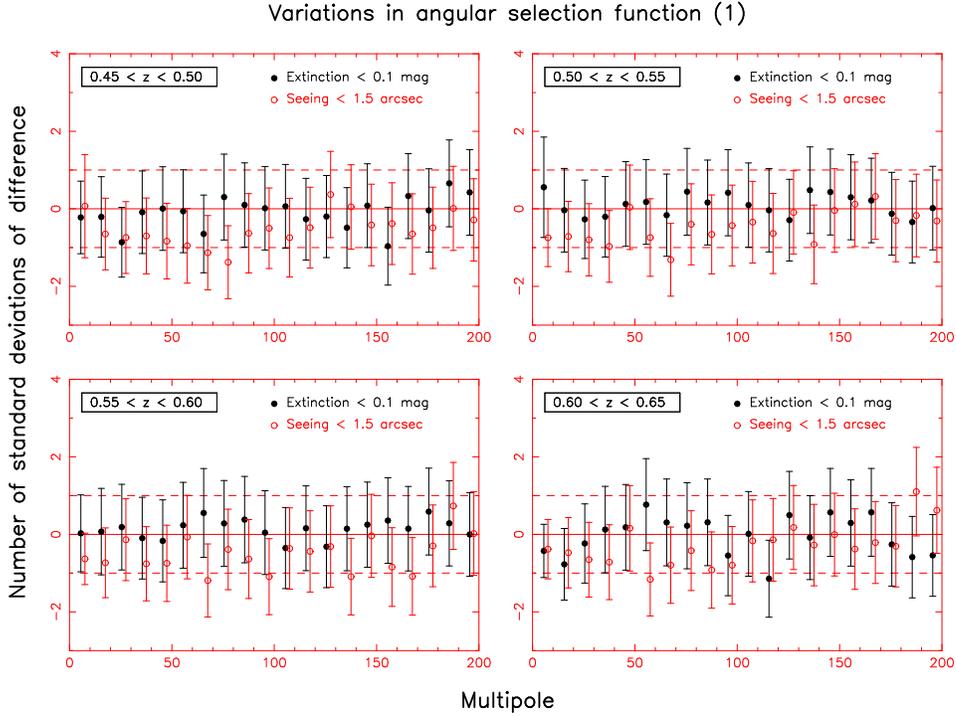}
\caption{The dependence of the angular power spectrum measurement in 4
  redshift slices on varying {\it dust extinction} and {\it
    astronomical seeing}.  Results from the default catalogue are
  compared to an analysis restricting the regions analyzed to (1) a
  maximum dust extinction of $0.1$ mag (see Figure \ref{figdust}), or
  (2) a maximum seeing of $1.5$ arcsec (see Figure \ref{figseeing}).
  The $x$-axis is multipole $\ell$ and the $y$-axis displays the
  number of standard deviations of the new measurement from the
  default result (with $\pm 1\sigma$ marked as horizontal dashed
  lines).}
\label{figsys1}
\end{figure*}

\begin{figure*}
\center
\epsfig{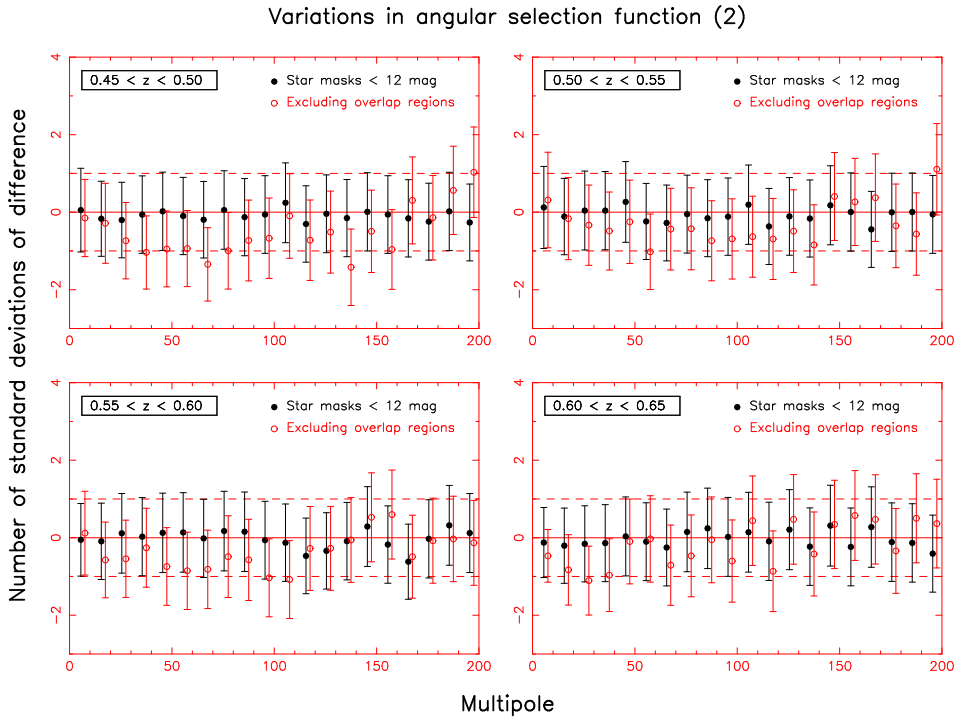}
\caption{The dependence of the angular power spectrum measurement in 4
  redshift slices on the presence of {\it bright objects} and {\it
    stripe overlap regions}.  Results from the default catalogue are
  compared to (1) an analysis placing circular masks of radius 1
  arcmin around all objects with $i$-band magnitudes brighter than 12,
  and (2) an analysis excluding overlap regions between stripes (see
  Figure \ref{figoverlap}).  The results are displayed in the same
  manner as Figure \ref{figsys1}.}
\label{figsys2}
\end{figure*}

\begin{figure*}
\center
\epsfig{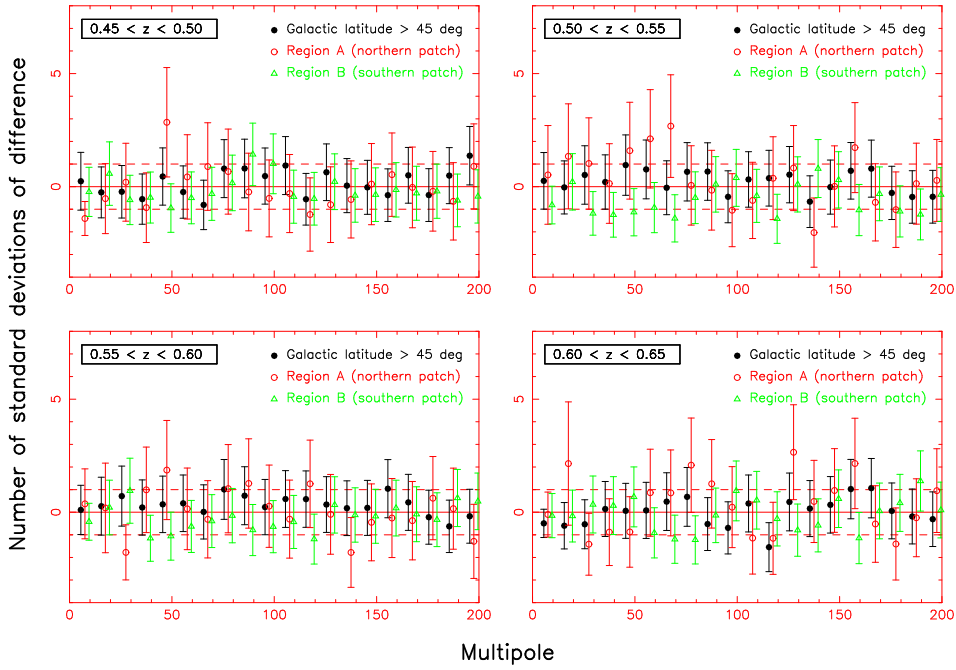}
\caption{The dependence of the angular power spectrum measurement in 4
  redshift slices on {\it Galactic latitude} and {\it disconnected
    region}.  Results from the default catalogue are compared to (1)
  an analysis excluding regions below Galactic latitude $45^\circ$,
  and (2,3) a separate analysis of each of the two largest
  disconnected regions in the survey window function (see Figure
  \ref{figwindow}).  The results are displayed in the same manner as
  Figure \ref{figsys1}.}
\label{figsys3}
\end{figure*}

\begin{figure*}
\center
\epsfig{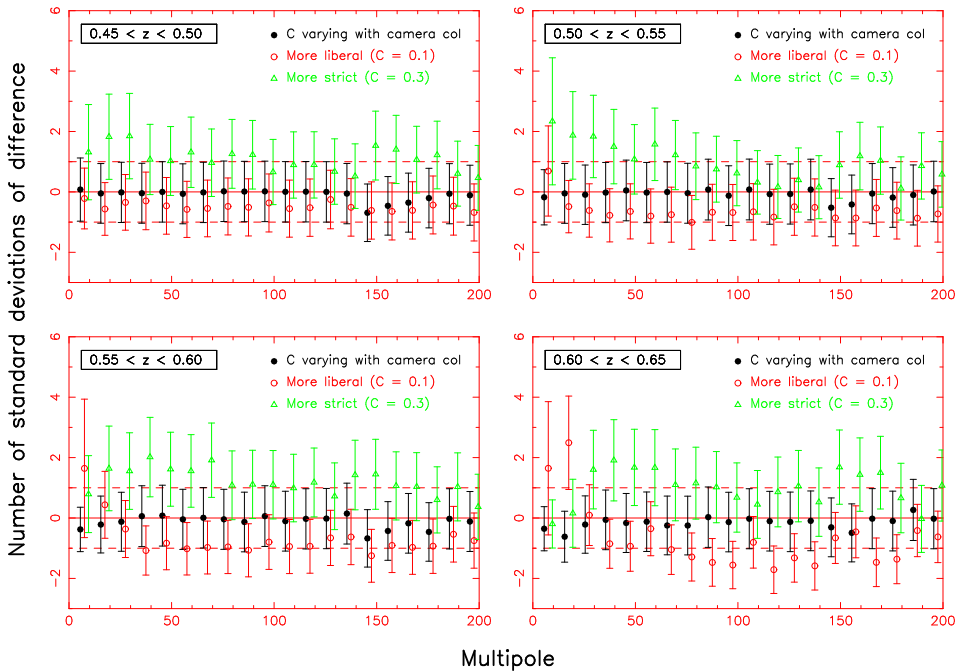}
\caption{The dependence of the angular power spectrum measurement in 4
  redshift slices on the star-galaxy separation criteria.  Results
  from the default catalogue are compared to analyses changing the
  value of the co-efficient $C$ in equation \ref{eqsgsep}.  The
  results are displayed in the same manner as Figure \ref{figsys1}.}
\label{figsys4}
\end{figure*}

\subsection{Dust extinction}

Figure \ref{figdust} displays the distribution of dust extinction
across the SDSS NGP (obtained from the public database).  Variations
in dust extinction may imprint systematic changes in galaxy density on
the sky near the magnitude limit of the survey, or if there are
systematic errors in the extinction map.  In this latter case, dust
extinction may affect photometric redshift estimates by reddening
galaxy colours.  We repeated the angular power spectrum measurement
excluding regions with high dust extinction ($> 0.1$ mag; $14\%$ of
the survey area).  The results in each of the four redshift slices are
shown by the solid circles in Figure \ref{figsys1}.  These data points
illustrate the number of standard deviations by which the angular
power spectrum changes from its default value given the variation in
the selection criteria.  Restricting the analyzed area to regions of
low dust extinction produces no significant change in the power
spectrum measurement.

\subsection{Astronomical seeing}

Figure \ref{figseeing} displays the distribution of seeing across the
SDSS NGP (obtained from the public database).  Poor seeing affects the
reliability of the galaxy photometry and the efficacy of the
star-galaxy separation.  We repeated the angular power spectrum
measurement excluding regions with poor seeing ($> 1.5$ arcsec; $15\%$
of the survey area).  The result is shown by the open circles in
Figure \ref{figsys1}.  Again, any change in the value of the power
spectrum is less than the statistical error in the measurement.

\subsection{Incompleteness near bright objects}

In the immediate vicinity of bright stars or galaxies, the depth
(completeness) of the SDSS imaging survey decreases owing to the
difficulty in distinguishing close neighbours to these bright objects
against the enhanced background.  The result is a spurious reduction
in the number of very close galaxy pairs (with separations $\theta <
0.02^\circ$) at faint magnitude thresholds.  This effect is negligible
for our analysis owing to the large angular scales we are considering
(multipoles $\ell \le 500$ of the angular power spectrum) and to our
exclusion of bright ($i_{\rm deV} < 17.5$) and faint ($i_{\rm deV} >
19.8$) objects.  In order to verify this fact, we repeated the angular
power spectrum measurement amending the angular selection function by
placing a series of circular masks of radius $1'$ around very bright
SDSS imaging objects with magnitudes $i < 12$ (removing $1.4\%$ of the
survey area).  The result is shown by the solid circles in Figure
\ref{figsys2}, which differ from the default measurement by much less
than the statistical error.

\subsection{Stripe overlap regions}

The SDSS is performed in scans which result in long data stripes of
width $2.5^\circ$ spanning the sky.  These stripes overlap towards the
edges of the surveyed region, as displayed in Figure \ref{figoverlap},
causing an increase in the effective depth of the survey in these
areas of overlap (or equivalently, an increase in the completeness at
a fixed magnitude threshold near the survey limit).  We repeated the
angular power spectrum measurement excluding these overlap regions
($18\%$ of the survey area).  The resulting deviation is shown by the
open circles in Figure \ref{figsys2}.  Any differences are small, and
we conclude that our faint magnitude threshold of $19.8$ is
sufficiently bright that differential completeness effects are
unimportant in our analysis.

\subsection{Galactic latitude}

The residual stellar contamination in our photometric database may be
a function of Galactic latitude, potentially imprinting large-scale
systematic density fluctuations.  Lines of constant Galactic latitude
are overlaid on the survey window function in Figure \ref{figwindow}.
We repeated the angular power spectrum measurement excluding areas
with Galactic latitudes below $45^\circ$ ($30\%$ of the survey area).
The resulting deviation is shown by the solid circles in Figure
\ref{figsys3}.  Again, the differences can be largely accounted for by
random statistical error, and we conclude that any sky variation of
stellar contamination has a negligible effect on the clustering
measurements.

\subsection{Analysis of disconnected regions}

The survey window function (Figure \ref{figwindow}) is mainly
comprised of two large disconnected regions.  These northern and
southern regions, which we will call A and B, comprise $32\%$ and
$67\%$ of the total survey area, respectively.  The lack of overlap
between these regions implies that relative photometric calibration is
challenging.  It is plausible that some systematic calibration
difference between these two regions could imprint apparent
large-scale fluctuations in power.  We repeated the angular power
spectrum measurement for each region separately.  The resulting
deviations are shown by the open circles and triangles in Figure
\ref{figsys3}, which reveal no systematic difference in results from
the joint analysis of the combination of regions.  We conclude that
the accuracy of the relative photometric calibration between these two
regions is acceptable.

\subsection{Star-galaxy separation}
\label{secsgsep}

The unavoidable presence of a small fraction $f$ of stars in our
analyzed catalogue affects the clustering measurement.  The addition
of a distribution of uncorrelated stars to our clustered galaxies
reduces the overall amplitude of the power spectrum.  In equation
\ref{eqclapprox}, the effect is that the redshift probability
distribution $p(z)$ is now normalized to $(1-f)$.  To first order, the
clustering amplitude is reduced by a factor $(1-f)^2$, which may be
conveniently absorbed into the constant galaxy bias factor fitted to
the data in equation \ref{eqclapprox}.

There are also second order effects.  Firstly, problems would arise at
low Galactic latitudes owing to the varying stellar density with
position.  This is not important for the surveyed SDSS regions, as
demonstrated above.  Secondly, the optical quality of the SDSS camera
varies a little with camera column (i.e., perpendicular to the stripe
direction) across its $2.5^\circ$ diameter.  This affects our ability
to separate stars and galaxies as a function of SDSS $\nu$-coordinate
(i.e., across stripes).  Near the edges of the camera, the
point-spread function is a little broader, and thus faint compact
galaxies will be preferentially mistaken for stars, inducing a
fluctuating galaxy surface density with minima at the edges of
stripes.  The angular scale of the resulting fluctuations will be
similar to the width of the stripes, $2.5^\circ$.  Unfortunately, this
is comparable to the preferred angular scale of the baryon
oscillations ($\approx 5^\circ$ at $z = 0.5$).

\begin{figure}
\center
\epsfig{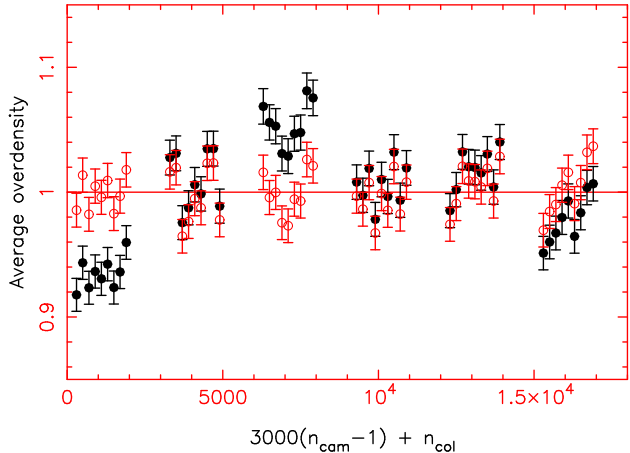}
\caption{The average overdensity of LRGs as a function of pixel
  position across the SDSS camera (perpendicular to the direction of
  the camera columns).  This position is quantified by the value of
  $3000 \times (n_{\rm cam} - 1) + n_{\rm col}$, where $n_{\rm cam}$
  is the camera column, $n_{\rm col}$ is the pixel position of the
  galaxy centroid in that column, and 3000 is an arbitrary offset to
  separate the camera columns clearly in the Figure.  Owing to the
  varying optical quality of the camera across the field-of-view, if
  constant star-galaxy separation criteria are applied then the
  resulting overdensity profile will exhibit a peak at the centre of
  the field with minima at the edges (as shown by the solid circles).
  In order to correct for this effect, we varied the aggressiveness of
  the star-galaxy separation as a function of camera column as
  described in the text, removing the systematic density variation (as
  shown by the open circles).}
\label{figsgcamcol}
\end{figure}

In Figure \ref{figsgcamcol} we illustrate this effect by plotting the
average overdensity of the LRG catalogue as a function of pixel
position across the field-of-view of the SDSS camera.  We bin galaxies
as a function of the quantity $x = 3000 \times (n_{\rm cam} - 1) +
n_{\rm col}$, where $n_{\rm cam}$ is the SDSS camera column containing
the object (an integer between 1 and 6), $n_{\rm col}$ is the pixel
position of the galaxy centroid in that column (a decimal up to 2048),
and 3000 is an arbitrary offset to separate the camera columns clearly
in the Figure.  The solid points represent the average galaxy
overdensity using the star-galaxy separation criteria described by
Collister et al.\ (2006),
\begin{eqnarray}
& i_{\rm psf} - i_{\rm model} \ge C \times (21 - i_{\rm deV})
  \nonumber \\
& \theta_{i,{\rm deV}} \ge 0.2''
\label{eqsgsep}
\end{eqnarray}
where the coefficient $C = 0.2$.  In this analysis we do {\it not}
include any additional neural network star-galaxy separation using
ANNz (see Section \ref{secdata}).  For the reasons discussed above,
the resulting LRG overdensity peaks in column 3 and has minima at the
camera edges (columns 1 and 6).  In order to correct for this
systematic imprint, we tried varying the coefficient $C$ in equation
\ref{eqsgsep} to trace the quality of the star-galaxy separation, such
that $C$ assumed different values $(0.1, 0.2, 0.25, 0.2, 0.2, 0.16)$
in the six camera columns.  These modified values of $C$ are chosen to
ensure that the average galaxy overdensity is constant across the
camera field (the open circles in Figure \ref{figsgcamcol}).

We then repeated the angular power spectrum measurement comparing our
catalogue with default star-galaxy separation ($C = 0.2$) to the model
where $C$ varies with camera column.  We also considered alternative
constant values $C = 0.1$ and $C = 0.3$.  The results are displayed in
Figure \ref{figsys4}.  Our first conclusion is that the adoption of
the camera column-dependent star-galaxy separation is of negligible
benefit.  The imprint of the varying overdensity in Figure
\ref{figsgcamcol} in fact has no measurable effect on the power
spectrum.  We therefore retain our default criteria of a constant
coefficient $C = 0.2$.  The effect of changing this value is in
reasonable accord with our model of a constant factor in power, which
may be safely absorbed into the fitted galaxy bias factor.  The
possible exception to this conclusion is the lowest multipole bands in
the furthest redshift slice $0.6 < z < 0.65$ (which is expected to
contain the most significant stellar contamination).  In this case,
more liberal star-galaxy separation actually increases the observed
power on the largest scales.  In fact, the lowest multipole band for
this redshift slice is observed to possess marginally significant
excess power in Figure \ref{figclpanels}, thus this one data point may
be influenced by stellar contamination.

\section{Measurement of the spatial power spectrum}
\label{secspatpow}

In this Section we determine the spatial galaxy power spectrum by a
direct three-dimensional Fourier transform of the gridded photo-$z$
catalogue.  We believe that this is a novel technique which has not
been previously applied to galaxy data, although Seo \& Eisenstein
(2003) and Blake \& Bridle (2005) give theoretical discussions.

\subsection{Method}

The photo-$z$ error distribution produces a convolution of the
underlying density field in the radial direction, and hence (in the
first approximation) a multiplicative radial damping of the measured
three-dimensional power in Fourier space.  In principle this damping
factor can be calculated using the known photo-$z$ error properties
and then divided out, restoring the underlying unconvolved power
spectrum.  We must implement a flat-sky approximation such that the
photo-$z$ smearing only occurs along one axis.

For example, in the ideal case that the radial co-moving co-ordinate
$x$ of each galaxy is smeared by an amount $\delta x$ sampled from a
Gaussian distribution
\begin{equation}
f(\delta x) \propto \exp{ \left[ -\frac{1}{2} \left( \frac{\delta
      x}{\sigma_x} \right)^2 \right] }
\label{eqgauss}
\end{equation}
then the resulting power spectrum signal will be damped in the radial
direction in accordance with:
\begin{equation}
P(k_x,k_y,k_z) \rightarrow P(k_x,k_y,k_z) \times \exp{ [-(k_x \,
    \sigma_x)^2]}
\label{eqpkdamp}
\end{equation}
Thus Fourier modes with high values of $|k_x| \gg 1/\sigma_x$
contribute only noise to the measurement.  However, if we only retain
power-spectrum modes with $|k_x| < k_{x,{\rm max}} \sim 1/\sigma_x$,
then we can deconvolve the measured 3D power spectrum by dividing by
the function $\exp{ [-(k_x \, \sigma_x)^2]}$ prior to combining the
Fourier modes in bins of fixed $k = \sqrt{k_x^2 + k_y^2 + k_z^2}$
(where $k_x < k_{x,{\rm max}}$).  In practice the photo-$z$ error
distribution is not a constant Gaussian function, and we calculated
the deconvolution function for our sample using Monte Carlo
simulations rather than an analytic approximation, as described below.

We initially analyzed the galaxies in photometric redshift slices of
width $\Delta z = 0.05$, as for the angular power spectrum
measurement.  We implemented the flat-sky approximation by dividing
the survey area into patches of width $20 \times 20$ deg.  These
patches have sufficient area to contain a reasonable number of
large-scale power spectrum modes with $k < 0.2 \, h$ Mpc$^{-1}$, and
yet are small enough that the photo-$z$ smearing may be considered to
act along one axis of a survey cuboid (which we take as the $x$-axis).
We only analyzed patches for which the survey angular selection
function covered over $50\%$ of the area of the patch.  The equations
mapping the galaxy right ascensions and declinations $(\alpha,\delta)$
onto spatial positions $(X,Y,Z)$ in the cuboid are
\begin{eqnarray}
X &=& \frac{dx}{dz}(z_0) \times (z - z_0) \nonumber \\
Y &=& x(z_0) \times (\alpha - \alpha_0) \cos{\delta_0} \nonumber \\
Z &=& x(z_0) \times (\delta - \delta_0)
\label{eq2dto3d}
\end{eqnarray}
where $(\alpha_0,\delta_0)$ are the central angular co-ordinates of
each patch, $z$ is the galaxy redshift, $z_0$ is the central redshift
of each slice, and $x(z)$ is the co-moving radial distance of the
survey redshift.  Thus we must assume values for the cosmological
parameters to construct this mapping, as discussed below.

We then measured the 3D power spectrum of galaxies in each cuboid in
Fourier bins of width $\Delta k = 0.01 \, h$ Mpc$^{-1}$, correcting
for survey window function effects and assuming a radial selection
function which we directly fitted to the survey data.  We only
included a Fourier mode $(k_x,k_y,k_z)$ in the analysis if the radial
wavescale $|k_x| < k_{x,{\rm max}} = 1.5/\sigma_x$, where $\sigma_x$
is the standard deviation of the photo-$z$ errors within the slice in
co-moving co-ordinates.  For redshift slices of thickness $\Delta z =
0.05$ this restricts us to modes with $k_x = 0$, because the spacing
of the modes in radial $k$-space is $\Delta k_x = 2\pi/L_x > k_{x,{\rm
    max}}$, where $L_x$ is the width of the slice in co-moving
co-ordinates (we also calculate the spatial power spectrum for a wider
redshift slice with $\Delta z = 0.15$).  The final power spectrum was
determined by averaging the measurements in the separate patches.  The
error in the power spectrum was taken to be the standard deviation of
the measurements in the patches.

The calculation of the spatial power spectrum $P(k)$ in this manner
depends upon the choice of cosmological parameters (which specifies
the cosmological distances $x(z)$ in equation \ref{eq2dto3d}).
Therefore, the determination of best-fitting cosmological parameters
using this method must proceed in a more complex fashion than the
angular power spectrum fits in Section \ref{secangpow}.  In a
measurement of $P(k)$, both the data and the model depend on the trial
cosmological parameters.  In contrast, the measured angular power
spectrum does not depend on the cosmological parameters and may be
left fixed during the model-fitting process.

However, there is an important advantage in performing a direct
measurement of $P(k)$ from a photo-$z$ catalogue: we avoid unnecessary
radial binning of the data.  The projection of $P(k)$ to the angular
power spectrum $C_\ell$ (equation \ref{eqclapprox}) corresponds to a
convolution of $P(k)$ in Fourier space in which the contrast of any
features (such as the baryon oscillations) will be reduced.  But
direct manipulation of the 3D power spectrum involves only a
multiplicative damping of power (equation \ref{eqpkdamp}) which does
not diminish the contrast of features.  Hence the direct power
spectrum technique presented in this Section may yield the most
significant detection of baryon oscillations.  In this initial study
we do not in fact fit cosmological parameters to the $P(k)$
measurements, but instead assume the best-fitting set of parameters
from the angular power spectrum analysis (Table \ref{tabpar}) in order
to demonstrate the consistency of the results and to search for the
existence of baryon oscillations.

\begin{figure*}
\center
\epsfig{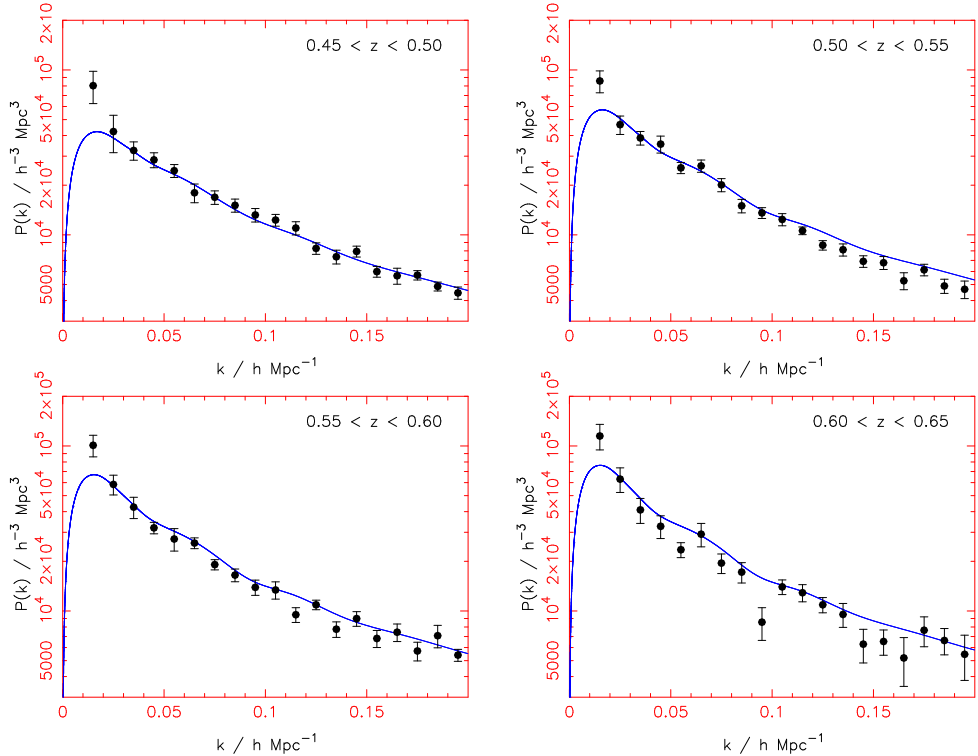}
\caption{The spatial power spectrum of the photo-$z$ catalogue in each
  of four redshift slices of width $\Delta z = 0.05$, measured using
  direct Fourier techniques in a series of $20 \times 20$ deg patches.
  There is no data point in the first bin (at $k = 0.005 \, h$
  Mpc$^{-1}$) because this bin contains no Fourier modes in the
  patches.  The solid lines are not a fit to this data, but are the
  model power spectra corresponding to the best-fitting cosmological
  parameters $(\Omega_{\rm m}, f_{\rm b}, \sigma_8, b)$ for the
  angular power spectrum analysis in each redshift slice.  The good
  agreement between the data and model validates our analysis
  techniques.}
\label{figpkpanels}
\end{figure*}

The spatial power spectrum measured by this direct Fourier calculation
must be carefully normalized when the condition $\sigma_x \ll L_x$ is
not satisfied, i.e., when the photo-$z$ error is comparable to the
width of the survey slice (which is the case here).  For a
spectroscopic survey of volume $V$, the dimensionless power spectrum
of the Fourier amplitudes, $|\delta_k|^2$, must be multiplied by $V$
in order to obtain a measured quantity (e.g.\ with units of $h^{-3}$
Mpc$^3$) which is independent of the volume of the survey and may be
compared with theoretical predictions.  Suppose this spectroscopic
survey is now convolved with a photo-$z$ error function, and a radial
slice of the resulting photo-$z$ catalogue is analyzed, with volume
$V_{\rm slice}$.  The correct normalization factor for comparison of
the measured power spectrum with theory is not in fact $V_{\rm
  slice}$, because the galaxy distribution within the slice has been
smoothed over a wider volume owing to the scattering of objects into
the slice by the photo-$z$ error distribution.  In the ideal case of a
``top-hat'' radial selection function for the photo-$z$ slice (of
width $L_x$) and the Gaussian photo-z error function of equation
\ref{eqgauss}, the multiplicative normalization correction is given by
\begin{equation}
N = \frac{1}{2\pi} \int_{-\infty}^\infty du \, \exp{ \left[ -\left(
    \frac{u}{L_x/\sigma_x} \right)^2 \right]} \left[
  \frac{\sin{(u/2)}}{(u/2)} \right]^2
\label{eqpknorm}
\end{equation}
such that the effective normalization volume for the power spectrum
measurement is $V_{\rm slice}/N$.  Equation \ref{eqpknorm} has the
limits $N \rightarrow 1$ for $\sigma_x \ll L_x$ (i.e., a spectroscopic
survey) and $N \rightarrow L_x/(2 \sqrt{\pi} \sigma_x)$ for $L_x \ll
\sigma_x$.

Moreover, the smoothing of the density distribution from outside the
photo-$z$ slice affects the shot noise correction term $P_{\rm shot}$
subtracted from the raw power spectrum measurement, $|\delta_k|^2$.
The shot noise correction is not simply given by the volume density
$n$ of analyzed galaxies within the slice, $P_{\rm shot} = 1/n$.  Thus
the measured power spectrum $P_{\rm meas}$ must be adjusted by both
additive (shot noise) and multiplicative (normalization) corrections
in order to achieve a reliable estimate of the underlying power
spectrum $P_{\rm true}$:
\begin{equation}
P_{\rm true}(k) = A(k) \, P_{\rm meas}(k) + B(k) \, P_{\rm shot}
\label{eqpkcorr}
\end{equation}
where $A(k)$ and $B(k)$ are roughly constant with $k = \sqrt{k_x^2 +
  k_y^2 + k_z^2}$, although change rapidly with $k_x$ ($\le k_{x,{\rm
    max}} \ll k)$ owing to the exponential damping factor in equation
\ref{eqpkdamp}.

We determined the two correction functions $A(k)$ and $B(k)$ in
equation \ref{eqpkcorr} using Monte Carlo realizations.  We first
generated clustered distributions of galaxies in a three-dimensional
survey cone, using a model spatial power spectrum corresponding to the
best-fitting cosmological parameters.  These Gaussian power-spectrum
realizations were created using the method described in Glazebrook \&
Blake (2005).  The radial selection function before the application of
photometric redshift errors was fixed by the spectroscopic redshift
distribution of the 2SLAQ training set (our best estimate of the
unconvolved redshift distribution).  The SDSS angular selection
function of Figure \ref{figwindow} was then imposed.  We then applied
a photometric redshift convolution function, again determined using
the training set, such that a different photo-$z$ error distribution
is implemented at each spectroscopic redshift, to ensure maximum
consistency with the statistics of the real catalogue.  The
appropriate photo-$z$ slices were then isolated and the spatial power
spectra were measured with the same methods as applied to the real
data.  In order to complete the calculation, a second set of Monte
Carlo realizations was generated using an identical method, except
that the original spectroscopic catalogue was unclustered (i.e.,
$P_{\rm true} = 0$ in equation \ref{eqpkcorr}) and thus the measured
power spectrum purely probes shot noise effects.  The two measured
power spectra, of the clustered and unclustered simulations, in
conjunction with the known input power spectrum, allow us to deduce
the functions $A(k)$ and $B(k)$ in equation \ref{eqpkcorr}.  These
correction factors are then applied to our power spectrum measurements
from the real data.

\subsection{Results}

We display our spatial power spectrum measurements in the four
redshift slices in Figure \ref{figpkpanels}, together with the models
corresponding to the best-fitting cosmological parameters from the
angular power spectrum analysis (Table \ref{tabpar}).  The agreement
is good, validating our analysis techniques.  There is evidence of
excess power in the largest-scale bin, possibly due to the failure of
the small-angle approximation of equation \ref{eq2dto3d} on these
scales.  We note that redshift-space distortions do not affect the
power spectrum models plotted in Figure \ref{figpkpanels} because we
are only analyzing power spectrum modes with $k_x = 0$.  The power
spectrum boost due to large-scale coherent infall depends on the angle
to the line-of-sight $\theta$ of a given Fourier mode as $(1 + \beta
\cos^2{\theta})$, thus modes with $k_x \approx 0$ (i.e. $\theta
\approx 90^\circ$) contain little imprint of redshift-space effects.

\begin{figure}
\center
\epsfig{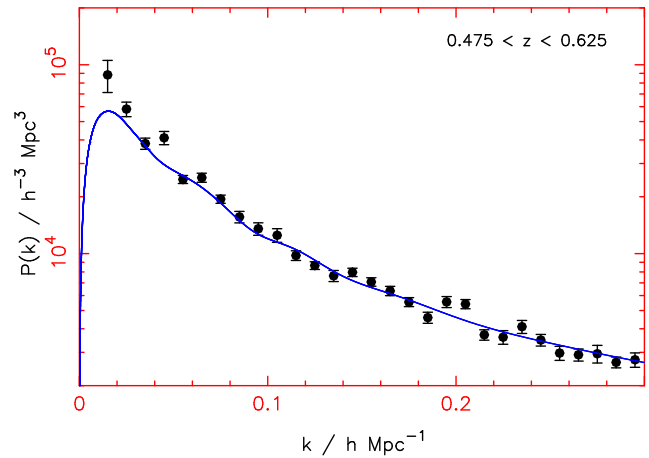}
\caption{The spatial power spectrum of the photo-$z$ catalogue in a
  wide redshift slice $\Delta z = 0.15$ such that Fourier modes with
  $k_x \ne 0$ are utilized in the analysis.  The agreement is good
  with the model power spectrum corresponding to the best-fitting
  cosmological parameters for the angular clustering measurements (the
  solid line).}
\label{figpktot}
\end{figure}

In Figure \ref{figpktot} we plot the results of analyzing a wider
redshift slice of the survey, $0.475 < z < 0.625$, such that Fourier
modes with $k_x \ne 0$ are being utilized.  Again, agreement with the
model spatial power spectrum is very good.  The shape of our recovered
power spectra also agree well with the measurement from the
spectroscopic sample of SDSS LRGs at lower redshift by Huetsi (2006a).

Unfortunately the current size of the survey is not sufficient to
yield convincing ``model-independent'' evidence for baryon
oscillations.  However, a modest expansion of the survey volume
(e.g.\ the final SDSS Data Release) would clearly create the potential
to produce such a detection (as modelled by Blake \& Bridle 2005).

\section{Conclusions}
\label{secconc}

We have quantified the patterns of large-scale structure in a
photometric-redshift catalogue of intermediate-redshift ($0.4 < z <
0.7$) Luminous Red Galaxies from the SDSS 4th Data Release.  This
catalogue, which we have named MegaZ-LRG, contains $> 10^6$
photometric redshifts derived with ANNz, an Artificial Neural Network
code (Collister et al.\ 2006).  In this study we analyze a
``conservative'' version of the catalogue, which includes $6 \times
10^5$ objects over $5{,}914$ deg$^2$.  The main conclusions of our
study are:

\begin{itemize}

\item We measured angular power spectra in a series of photo-$z$
  slices of width $\Delta z = 0.05$ from $z = 0.45$ to $z = 0.65$.
  The errors in the $C_\ell$ spectra were tested with Monte Carlo
  realizations and with a ``bootstrap'' analysis in sub-fields.  A
  simple Gaussian approximation was determined to be an accurate
  approximation to the error in $C_\ell$ over the scales of interest.

\item The angular power spectra are well-fitted by model spatial power
  spectra with consistent cosmological parameters $\Omega_{\rm m}
  \approx 0.27$, $f_{\rm b} \approx 0.17$, $H_0 = 75$ km s$^{-1}$
  Mpc$^{-1}$ and $n_{\rm scalar} = 1$ over the series of redshift
  slices.

\item Combining the results for the different redshift slices with
  appropriate covariances results in a measurement of $\Omega_{\rm m}
  h = 0.195 \pm 0.023$ and $\Omega_{\rm b}/\Omega_{\rm m} = 0.16 \pm
  0.036$ (for fixed values of $H_0$ and $n_{\rm scalar}$,
  marginalizing over $\sigma_8$ and $b$).  These measurements agree
  with the latest analysis of the Cosmic Microwave Background
  radiation (the 3-year results of the Wilkinson Microwave Anisotropy
  Probe, see Spergel et al.\ 2006).  The measurements from the CMB
  data alone are $\Omega_{\rm m} h = 0.17 \pm 0.03$ and $\Omega_{\rm
    b}/\Omega_{\rm m} = 0.18 \pm 0.01$.

\item These results agree very well with a parallel investigation of a
  similar dataset carried out by Padmanabhan et al.\ (2006) using a
  different photometric-redshift estimation method and power spectrum
  analysis procedure.  Padmananhan et al.\ quote cosmological
  parameter fits $\Omega_{\rm m} h = 0.21 \pm 0.021$ and $\Omega_{\rm
    b}/\Omega_{\rm m} = 0.18 \pm 0.04$, in both cases within
  1-$\sigma$ of our results.

\item We detect visual suggestions of baryon oscillations with a
  statistical significance of less than 3-$\sigma$.

\item We performed a direct re-construction of the spatial power
  spectrum using a Fourier analysis which only retains large-scale
  radial components.  We demonstrated that this method produces
  results which are consistent with the best-fitting model for the
  angular power spectrum analysis.

\item We searched for evidence of systematic photometric errors in the
  catalogue using a variety of tests, and were unable to identify any
  significant effects.

\item We measure a hint of excess power relative to the best-fitting
  cosmological model on the largest scales (the lowest multipole bands
  in the four redshift slices in Figure \ref{figclpanels}).  If
  confirmed, this excess power has a range of possible causes: (1)
  residual systematic errors; (2) cosmic variance; (3) large-scale
  galaxy biasing mechanisms; (4) new early-Universe physics.

\end{itemize}

The cosmological parameter measurements we obtain from the large-scale
clustering of this photometric redshift catalogue have a similar
precision to those derived from the latest spectroscopic redshift
surveys, despite the weaker redshift information.  For example, Cole
et al.\ (2005) report $\Omega_{\rm m} h = 0.17 \pm 0.02$ and
$\Omega_{\rm b}/\Omega_{\rm m} = 0.19 \pm 0.05$ from an analysis of
the final 2dF Galaxy Redshift Survey, fixing the Hubble parameter $h$
and scalar index of primordial fluctuations $n_{\rm scalar}$ in a
manner similar to our analysis.  We therefore conclude that
photometric redshift surveys have become competitive with
spectroscopic surveys for the measurement of cosmological parameters
using the large-scale clustering pattern in the simple ``vanilla''
model.

\section*{Acknowledgments}

We thank Daniel Eisenstein for useful discussions on a range of
topics.  We also acknowledge helpful conversations with Karl
Glazebrook, Ivan Baldry, Bob Nichol, Filipe Abdalla, David Schlegel
and Antony Lewis.  CB acknowledges support from the Izaak Walton
Killam Memorial Fund for Advanced Studies and from the Canadian
Institute for Theoretical Astrophysics National Fellowship programme.
OL acknowledges a PPARC Senior Research Fellowship.

Funding for the Sloan Digital Sky Survey (SDSS) has been provided by
the Alfred P. Sloan Foundation, the Participating Institutions, the
National Aeronautics and Space Administration, the National Science
Foundation, the U.S. Department of Energy, the Japanese
Monbukagakusho, and the Max Planck Society. The SDSS Web site is
{\tt http://www.sdss.org/}.

The SDSS is managed by the Astrophysical Research Consortium (ARC) for
the Participating Institutions. The Participating Institutions are The
University of Chicago, Fermilab, the Institute for Advanced Study, the
Japan Participation Group, The Johns Hopkins University, the Korean
Scientist Group, Los Alamos National Laboratory, the
Max-Planck-Institute for Astronomy (MPIA), the Max-Planck-Institute
for Astrophysics (MPA), New Mexico State University, University of
Pittsburgh, University of Portsmouth, Princeton University, the United
States Naval Observatory, and the University of Washington.

\end{document}